\documentclass[12pt]{elsarticle}

\usepackage{amssymb,amsmath,amsthm}

\usepackage{lineno}
\usepackage{multicol}
\usepackage{enumitem}
\usepackage{multirow}
\usepackage{mathtools}
\usepackage{algorithmic}
\usepackage{algorithm}
\usepackage{float}
\usepackage{braket}
\usepackage{graphicx}\graphicspath{{img/}}
\usepackage{dcolumn}\usepackage{bm}\usepackage{inputenc}
\usepackage{blindtext}
\usepackage{color}
\usepackage{booktabs,siunitx}
\usepackage{hyperref}\hypersetup{colorlinks,citecolor=blue}
\usepackage[svgnames]{xcolor}
\usepackage{listings}
\usepackage{comment}

\newcounter{bla}

\definecolor{codegreen}{rgb}{0.04313,0.55294,0}
\definecolor{codepurp}{rgb}{0.9294117,0.03529411,0.3411764}
\definecolor{codeblue}{rgb}{0,0.3450980,1}
\definecolor{codegray}{rgb}{0.1,0.11,0.1}
\definecolor{backcolour}{rgb}{0.97,0.97,0.97}

\lstdefinestyle{pystyle}{
    backgroundcolor=\color{backcolour},   
    commentstyle=\color{codeblue},
    keywordstyle=\color{codepurp},
    numberstyle=\tiny\color{codegray},
    stringstyle=\color{codegreen},
    basicstyle=\footnotesize,
    breakatwhitespace=false,         
    breaklines=true,                 
    captionpos=b,                    
    keepspaces=true,                 
    numbers=left,                    
    numbersep=5pt,                  
    showspaces=false,                
    showstringspaces=false,
    showtabs=false,                  
    tabsize=2
}
\journal{Computer Physics Communications}

\begin{document}

\newcommand{\mean}[1]{\big\langle#1\big\rangle}
\newcommand{\dint}{\ensuremath{\mathrm{d}}}
\newcommand{\rvec}{\ensuremath{\mathbf{r}}}
\newcommand{\Rvec}{\ensuremath{\mathbf{R}}}
\newcommand{\epsilonvec}{\ensuremath{\pmb\epsilon}}
\newcommand{\etavec}{\ensuremath{\pmb\eta}}
\newcommand{\sigmavec}{\ensuremath{\pmb\sigma}}
\newcommand{\uvec}{\ensuremath{\mathbf{u}}}
\newcommand{\Pvec}{\ensuremath{\mathbf{P}}}
\newcommand{\Fvec}{\ensuremath{\mathbf{F}}}
\newcommand{\Wvec}{\ensuremath{\mathbf{W}}}
\newcommand{\Xvec}{\ensuremath{\mathbf{X}}}
\newcommand{\Yvec}{\ensuremath{\mathbf{Y}}}
\newcommand{\Xd}{\ensuremath{\widetilde{D}}}
\newcommand{\Xmlip}{\ensuremath{\widetilde{\pmb{D}}}}
\newcommand{\gammavec}{\ensuremath{\pmb{\gamma}}}
\newcommand{\alphavec}{\ensuremath{\pmb{\alpha}}}
\newcommand{\Vmlip}{\ensuremath{\widetilde{V}_{\gammavec}}}
\newcommand{\qmlip}{\ensuremath{\widetilde{q}_{\gammavec}}}
\newcommand{\Nat}{\ensuremath{N_{\text{at}}}}
\newcommand{\Thetavec}{\ensuremath{\mathbf{\Theta}}}
\newcommand{\Zpart}{\ensuremath{\mathcal{Z}}}
\newcommand{\Fe}{\ensuremath{\mathcal{F}}}
\newcommand{\FeGB}{\ensuremath{\widetilde{\Fe}_{\gammavec}}}
\newcommand{\FeTDEP}{\ensuremath{\widetilde{\Fe}_{TDEP}}}
\newcommand{\Ra}{\ensuremath{\mathbfcal{R}}}
\newcommand{\He}{\ensuremath{\widetilde{\mathrm{H}}}}
\newcommand{\Ha}{\ensuremath{\mathrm{H}}}
\newcommand{\kBT}{\ensuremath{\mathrm{k}_\mathrm{B}\mathrm{T}}}
\newcommand{\epsi}{\ensuremath{\pmb{\varepsilon}}}
\newcommand{\DF}{\ensuremath{\mathcal{D}_{\mathrm{F}}}}
\newcommand{\DKL}{\ensuremath{\mathcal{D}_{\mathrm{KL}}}}
\newcommand{\FB}[1]{{\color{red} \bf #1}}
\newcommand{\PR}[1]{{\color{teal} \bf #1}}
\newcommand{\RB}[1]{{\color{cyan} \bf #1}}
\newcommand{\AC}[1]{{\color{lime} \bf #1}}
\newcommand{\CD}[1]{\textcolor{orange}{\bf [CD:] {#1}}}
\newcommand{\GA}[1]{{\color{Magenta} \bf #1}}
\begin{frontmatter}

\title{Machine learning assisted canonical sampling (\textsc{Mlacs})}

\address{}

\author[a,b,c]{Alo\"is Castellano}
\author[a,b]{Romuald B\'ejaud}
\author[a,b]{Pauline Richard}
\author[a,b,e]{Olivier Nadeau}
\author[a,b]{Cl\'ement Duval}
\author[a,b]{Gr\'egory Geneste}
\author[e]{Gabriel Antonius}
\author[c]{Johann Bouchet}
\author[f]{Antoine Levitt}
\author[g,h]{Gabriel Stoltz}
\author[a,b]{Fran\c{c}ois Bottin\corref{author}}

\cortext[author] {Corresponding author.\\\textit{E-mail address:} francois.bottin@cea.fr}
\address[a]{CEA, DAM, DIF, F-91297 Arpajon, France.}
\address[b]{Universit\'e Paris-Saclay, CEA, Laboratoire Mati\`ere en Conditions Extr\^emes, 91680 Bruy\`eres-le-Ch\^atel, France.}
\address[c]{NanoMat/Q-Mat/CESAM and European Theoretical Spectroscopy Facility, Université de Liège Belgium.}
\address[d]{CEA, DES, IRESNE, DEC, Cadarache, F-13018 St Paul Les Durance, France.}
\address[e]{Institut de recherche sur l’hydrogène, Université du Québec à Trois-Rivières, C.P. 500, Trois-Rivières, Canada G9A 5H7.}
\address[f]{Laboratoire de mathématiques d'Orsay, Université Paris-Saclay, France}
\address[g]{CERMICS, Ecole des Ponts, IP Paris, Marne-la-Vallée, France}
\address[h]{MATHERIALS team-project, Inria Paris, France}

\begin{abstract} 

The acceleration of material property calculations while maintaining {\it ab initio} accuracy (1 meV/atom) is one of the major challenges in computational physics. In this paper, we introduce a Python package enhancing the computation of (finite temperature) material properties at the {\it ab initio} level using machine learning interatomic potentials (MLIP). The Machine-Learning Assisted Canonical Sampling (\textsc{Mlacs}) method, grounded in a self-consistent variational approach, iteratively trains a MLIP using an active learning strategy in order to significantly reduce the computational cost of {\it ab initio} simulations.

\textsc{Mlacs} offers a modular and user-friendly interface that seamlessly integrates Density Functional Theory (DFT) codes, MLIP potentials, and molecular dynamics packages, enabling a wide range of applications, while maintaining a near-DFT accuracy.
These include sampling the canonical ensemble of a system, performing free energy calculations, transition path sampling, and geometry optimization, all by utilizing surrogate MLIP potentials, in place of {\it ab initio} calculations.

This paper provides a comprehensive overview of the theoretical foundations and implementation of the \textsc{Mlacs} method. We also demonstrate its accuracy and efficiency through various examples, showcasing the capabilities of the \textsc{Mlacs} package.

\end{abstract}


\end{frontmatter}

 {\bf PROGRAM SUMMARY}

\begin{small}
\noindent
{\em Program Title:} \textsc{Mlacs}                                         \\
{\em Licensing provisions:} GNU General Public License, version 3       \\
{\em Programming language:}  Python \\
{\em Journal reference of previous version:} The seminal version is defined in~\cite{Castellano_PRB106_2022}. \\
{\em Does the new version supersede the previous version?:} Yes. \\
{\em Reasons for the new version:} The new version~\cite{Mlacs_github}, \textsc{Mlacs} v1.0.1, works on various architectures and includes several new features.\\
{\em Nature of problem:}\\
 Numerous material properties, whether related to the ground state or finite temperature thermodynamic quantities, cannot be deduced from classical simulations and require accurate but highly demanding {\it ab initio} calculations. Enhancing the efficiency of these simulations while preserving a near-{\it ab initio} accuracy is one of the biggest challenges in modern computational physics.\\
{\em Solution method:}\\
 The emergence of MLIP potentials enables us to tackle this issue. The method implemented in \textsc{Mlacs} allows for the acceleration of {\it ab initio} calculations by training a MLIP potential on the fly. At the end of the simulation, \textsc{Mlacs} produces an optimal local surrogate potential, a database that includes a sample of representative atomic configurations with their statistical weights, as well as information on convergence control and thermodynamic quantities. \\

\end{small}

\section*{Introduction}
In contemporary materials science, Molecular Dynamics (MD) simulations, alongside Monte Carlo calculations, stand as prominent methods for attaining finite temperature properties and exploring materials phase diagrams. Notably, the pioneering works of Alder~\cite{Alder_JCP27_1957,Alder_JCP31_1959,Alder_JCP33_1960}, Rahman~\cite{Rahman_PR136_1964,Rahman_JCP55_1971} and collaborators established the groundwork for classical potentials, which subsequently paved the way for advancements in \textit{ab initio} molecular dynamics (AIMD)~\cite{Kresse_PRB47_1993,Kresse_PRB48_1993}. However, the computational demands inherent in AIMD, attributed to the evaluation of Hellmann--Feynman forces using density functional theory (DFT)~\cite{Hohenberg_PR136_1964,Kohn_PR140_1965} and the substantial number of MD time steps needed to sample the Born--Oppenheimer (BO)~\cite{Born_Annalen20_1927} surface in the canonical ensemble, impose limitations on the scope of systems that can be studied.

To tackle these computational challenges, two primary strategies have emerged. The first strategy circumvents AIMD simulations by directly generating atomic configurations to sample the canonical distribution, thus avoiding the computational overhead associated with AIMD simulations. A longstanding approach to derive temperature-dependent properties without resorting to molecular dynamics simulations is based on the harmonic approximation and its extension, such as
the Self-Consistent Harmonic Approximation (SCHA)~\cite{Gillis1968,Werthamer1970,Tadano2018,Esfarjani2020} 
which aim to generate configurations that best approximate the DFT canonical distribution. 
Among those, we can cite the stochastic Temperature Dependent Effective Potential~\cite{Shulumba_PRB95_2017}, the Stochastic SCHA~\cite{Bianco2017,Monacelli2018,Monacelli2021} and the Quantum Self-Consistent Ab Initio Lattice Dynamics~\cite{vanRoekeghem2021}. 
These methods, collectively referred to as Effective Harmonic Canonical Sampling (EHCS), generate configurations with displacements around equilibrium positions according to a distribution corresponding to an effective harmonic Hamiltonian, allowing to compute properties using \textit{ab initio} data. 
The self-consistent construction of this Hamiltonian, based on a variational procedure, allows for the inclusion of explicit temperature effects. 
However, the inherent Gaussian distribution of atomic displacements in these approaches may lead to discrepancies, particularly in highly anharmonic systems or near the melting temperature~\cite{Bouchet_PRB92_2015, Bouchet2017, Bouchet_PRB99_2019, Anzellini_2020, Castellano2020, Bouchet_JPCM34_2022, Bottin_PRB109_2024}, rendering them unsuitable for liquid-state simulations for example.

The second strategy involves replacing AIMD simulations with MD, using numerical potentials as surrogates to sample the ab initio BO surface.
Recent advancements in Machine Learning Interatomic Potentials (MLIP)~\cite{Zuo2020, Behler2016, Bartk2010, Thompson2015, Novikov2021} offer a promising avenue for accelerating finite-temperature studies with near-DFT accuracy. 
Their flexibility enables an accurate reproduction of diverse BO surfaces, making them a key tool for the study of the effects of temperature on material properties.
Yet, their construction remains a meticulous task, often necessitating careful selection and curation of training datasets.
Moreover, despite the impressive accuracy of MLIP, their use as surrogate can still lead to discrepancies compared to AIMD simulations due to accuracy or an evolution of the BO with the electronic temperature. In cases where stringent accuracy is needed, particularly for free energy problems, corrective schemes may be necessary~\cite{Liu_PRM_2021,Zhou2022}. 

In this work, we present the theory, the implementation, and the last developments of a recent method named Machine Learning Assisted Canonical Sampling (\textsc{Mlacs})\cite{Castellano_PRB106_2022}, which can be considered as a generalization of the EHCS methods to linear MLIP.
\textsc{Mlacs} employs a self-consistent variational procedure to generate configurations that best approximate the DFT canonical distribution, achieving near-DFT accuracy in computed properties at a fraction of the cost of AIMD. The MLIP in \textsc{Mlacs} is built to produce an effective canonical distribution that reproduces the DFT one at a single thermodynamical point, rather than approximating an effective BO surface for all thermodynamic conditions. Thus, \textsc{Mlacs} is a sampling method that accelerates the computation of finite-temperature properties compared to AIMD simulations, keeping the \textit{ab initio} system as the simulated system.

As we will demonstrate, \textsc{Mlacs} is a highly versatile tool that can accelerate AIMD, provide accurate total energies and forces useful for geometry relaxation, be used to find the minimum (free) energy path (MEP) between two atomic configurations or be combined with thermodynamic integration (TI) to obtain free energies and build phase diagrams. To date, \textsc{Mlacs} employs several kinds of linear MLIPs and can be extended to others thanks to its user-friendly ecosystem.

In the first section, we prove that the \textsc{Mlacs} approach is mathematically founded and show how the active learning strategy which results from the equations can be implemented. In a second section we extent this framework to thermodynamic integration, leveraging the optimal MLIP to accelerate the calculation of free energy at the {\it ab initio} level. In the third section we describe how this method can be used to go beyond canonical sampling and can address the geometry optimization issue or can strongly accelerate the search of minimum (free) energy paths. In a fourth section, we present the ecosystem of the \textsc{Mlacs} package, the python environment, the softwares and the MLIPs in the various workflows. Finally, in the fifth section, we provide several applications demonstrating the efficiency of \textsc{Mlacs} in challenging examples.

\section{{\it Ab initio} Canonical Sampling based on Variational Inference \label{sec:mlacs}}
\subsection{Theory}

In this section, we derive the formalism justifying the \textsc{Mlacs} approach~\cite{Castellano_PRB106_2022} and mathematically prove that a self-consistent active learning strategy using a MLIP enables to sample the BO surface  as accurately as using AIMD simulations.

Let us consider an arbitrary system of $\Nat$ atoms (a crystal or a liquid, elemental or alloyed) at a temperature $T$ (with $\beta = 1/\mathrm{k_B T}$) and described by the coordinates $\Rvec \in \mathbb{R}^{3 \Nat}$. For this system, the average of an observable $O(\Rvec)$, depending only on positions, is given by 
\begin{equation}
\label{eq:average}
    \braket{O} = \int \dint\Rvec O(\Rvec) p(\Rvec) 
               \approx \sum_n O(\Rvec_n) w_n \mbox{\; ,}
\end{equation}
with $w_n$ the weight of a configuration $n$,
which follows the normalization $\sum_n w_n = 1$, $p(\Rvec) = \mathrm{e}^{-\beta V(\Rvec)}/\Zpart$ the canonical equilibrium distribution induced by the BO potential surface $V(\Rvec)$ and $\Zpart=\int \dint\Rvec \mathrm{e}^{-\beta V(\Rvec)}$ the configurational canonical partition function. 

In the context of {\it ab initio} calculations, obtaining the canonical equilibrium distribution $p(\Rvec)$ often entails laborious AIMD simulations, which can be challenging to perform and may remain beyond reach, even when utilizing supercomputers with thousands of processors over extended periods, sometimes spanning months. This major disadvantage hinders the production of data with large supercells and long trajectories, limits the ability to conduct calculations for various thermodynamic conditions or systems, sometimes compromises the accuracy of the calculations and restricts the incorporation of more intricate physical considerations
into {\it ab initio} simulations. Instead of conducting AIMD simulations to generate a large set of $\Rvec_n$ and $w_n$, the aim of \textsc{Mlacs} is to generate a reduced and representative set of atomic configurations and weights using a MLIP, and subsequently compute them using DFT.

For this purpose we define, on one hand, a ``target" (DFT) system corresponding to a full AIMD equilibrated trajectory and, on the other hand, a ``surrogate" (MLIP) system corresponding to a sample of atomic configurations which are representative of the equilibrium canonical distribution. In the following, we will prove and demonstrate that the surrogate MLIP system will become closer to the DFT one as the iterations of the \textsc{Mlacs} algorithm proceed, reducing by several orders of magnitude the high computational cost of the target DFT system, while maintaining an {\it ab initio} accuracy. Let us define a MLIP linear potential 
\begin{equation}
\Vmlip(\Rvec) = \sum_{k=1}^K \Xd_k(\Rvec) \gamma_k \mbox{\; ,}
\end{equation}
with associated partition function $\widetilde{\Zpart}_{\gammavec}=\int \dint\Rvec \mathrm{e}^{-\beta \Vmlip(\Rvec)}$ and canonical equilibrium distribution $\qmlip(\Rvec)= \mathrm{e}^{-\beta \Vmlip(\Rvec)} / \widetilde{\Zpart}_{\gammavec}$. The column vector $\gammavec \in \mathbb{R}^K$ contains adjustable parameters and the functions included in the row vector $\Xmlip: \mathbb{R}^{3\Nat}\rightarrow \mathbb{R}^K$ are named descriptors. All the equilibrium quantities of the ``target" and ``surrogate" systems computed in the canonical ensemble and introduced in this section are listed in Table~\ref{tab:def}.

\begin{table*}
\caption{\label{tab:def} Definition of various equilibrium quantities of the target and surrogate systems in the canonical ensemble.}
\begin{tabular}{crlrl}
\hline \hline
System & \multicolumn{2}{l}{Target (DFT)} & \multicolumn{2}{l}{Surrogate (MLIP)} \\ 
\hline \\
 Potential energy & $V(\Rvec)$ & & $\Vmlip(\Rvec)$ & $= \Xmlip(\Rvec) \gammavec$ \\ 
 & & & & $= \sum_{k=1}^K \Xd_k(\Rvec) \gamma_k$ \\
 Distribution & $p(\Rvec)$ & $= \frac{\mathrm{e}^{-\beta V(\Rvec)}}{\Zpart}$ & $\qmlip(\Rvec)$ & $= \frac{\mathrm{e}^{-\beta \Vmlip(\Rvec)}}{\widetilde{\Zpart}_{\gammavec}}$ \\
\begin{tabular}{@{}c@{}}Configurational \\ Partition Function\end{tabular} & $\Zpart$ & $=\int \text{d}\Rvec \mathrm{e}^{-\beta V(\Rvec)}$ & $\widetilde{\Zpart}_{\gammavec}$ & $=\int \text{d}\Rvec \mathrm{e}^{-\beta \Vmlip(\Rvec)}$ \\
 Free energy & $\Fe $ & $= -\kBT \ln(\Zpart)$ & $\FeGB^0 $ & $= -\kBT \ln(\widetilde{\Zpart}_{\gammavec})$ \\
 Average & $\braket{O} $ & $= \int\dint\Rvec O(\Rvec) p(\Rvec)$ & $\braket{O}_{\Vmlip} $ & $= \int \dint\Rvec O(\Rvec) \qmlip(\Rvec)$ \\
\hline \hline
\end{tabular}
\end{table*}

\begin{figure}
    \centering
    \includegraphics[width=0.8\linewidth]{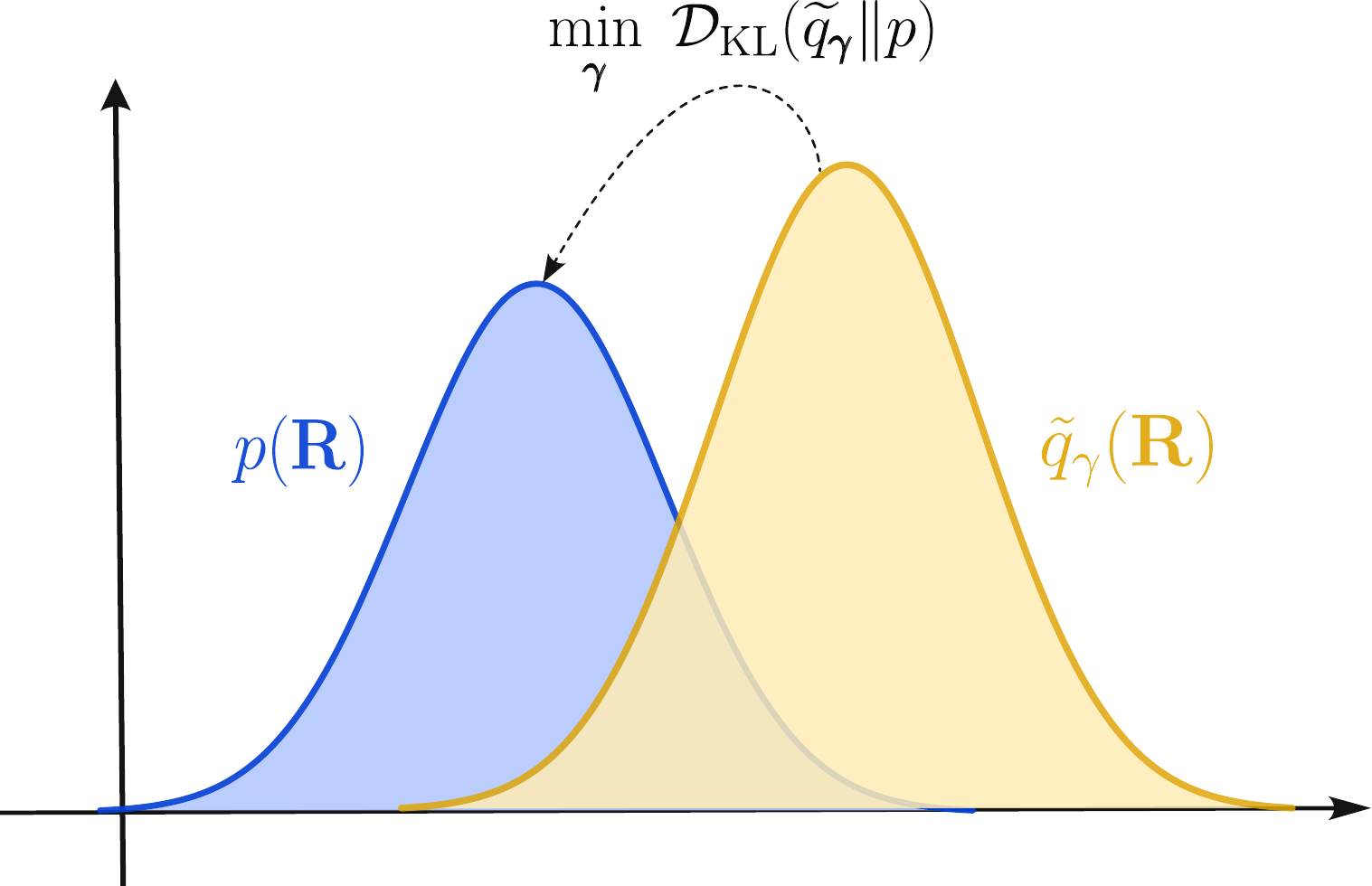}
    \caption{Measure of the similarity between two distributions $p(\Rvec)$ and $\qmlip(\Rvec)$ based on the Kullback--Leibler divergence $\DKL(\qmlip\Vert p)$.}
    \label{fig:kld}
\end{figure}

A pivotal concept in \textsc{Mlacs} involves utilizing the distribution $\qmlip(\Rvec)$ instead of $p(\Rvec)$ in Eq.~\eqref{eq:average}. This strategy leverages the accelerated computational efficiency gained by generating configurations using $\Vmlip(\Rvec)$ rather than $V(\Rvec)$. To ensure precision in the results, one must adjust the parameters $\gammavec$ of the surrogate potential such that $\qmlip(\Rvec)$ serves as the optimal approximation to the true distribution $p(\Rvec)$. The Kullback--Leibler divergence (KLD) 
\begin{equation}
\label{eq:KL}
    \DKL(\qmlip\Vert p) = \int \dint\Rvec \qmlip(\Rvec) \ln \left(\frac{\qmlip(\Rvec)}{p(\Rvec)}\right) \geq 0 \mbox{\; ,}
\end{equation}
based on the Shannon entropy, is a non-negative and asymmetric measure of the discrepancy between the two distributions $p$ and $\qmlip$. The smaller the KLD, the closer the distributions $p$ and $\qmlip$ are, with $\DKL(\qmlip\Vert p)=0$ indicating identical distributions. Therefore, by minimizing this quantity with respect to the parameters $\gammavec$, we obtain the distribution $\qmlip(\Rvec)$ that best approximates $p(\Rvec)$ (see Fig.~\ref{fig:kld}). 

A more manageable formulation can be achieved by replacing $\qmlip(\Rvec)$ and $p(\Rvec)$ by their expression in Eq.~\eqref{eq:KL}. The problem then transforms into an equivalent free energy minimization:
\begin{equation}
\begin{split}
        \DKL(\qmlip \Vert p) &= \int \dint \Rvec \qmlip(\Rvec) \bigg( \ln \bigg[\frac{\Zpart}{\widetilde{\Zpart}_{\gammavec}}\bigg] + \ln \bigg[ \mathrm{e}^{-\beta \big( \Vmlip(\Rvec) - V(\Rvec) \big)} \bigg] \bigg) \\
                             &= \ln (\Zpart) - \ln(\widetilde{\Zpart}_{\gammavec}) + \beta \braket{V(\Rvec) - \Vmlip(\Rvec)}_{\Vmlip} \\
                             &=-\beta (\Fe -\FeGB)  \mbox{\; ,}
\label{eq:DKL_GB}
\end{split}
\end{equation}
where $\braket{O}_{\Vmlip} = \int \dint \Rvec O(\Rvec) \qmlip(\Rvec)$ is the canonical average for the potential $\Vmlip(\Rvec)$, $\Fe = -\mathrm{k_B T} \ln(\Zpart)$ the free energy associated with $V(\Rvec)$, $\FeGB^0 = -\mathrm{k_B T} \ln(\widetilde{\Zpart}_{\gammavec})$ the free energy associated with $\Vmlip(\Rvec)$ and 
\begin{equation}
\label{eq:eqGB}
\FeGB = \FeGB^0 + \braket{V(\Rvec) - \Vmlip(\Rvec)}_{\Vmlip} \mbox{\; ,}
\end{equation}
the Gibbs--Bogoliubov free energy.
Thus, the minimization of the KL divergence (see Eq.~\eqref{eq:KL}) is equivalent to the minimization of the Gibbs--Bogoliubov inequality: 
\begin{equation}
\label{eq:ineqGB}
    \Fe           \leq \min_{\gammavec} \; \FeGB \mbox{\; ,}
\end{equation}
which states that the free energy of the target system is always lower than the Gibbs--Bogoliubov free energy calculated for any trial potential~\Vmlip. This inequality is the starting point for various variational procedures, in particular the Self-Consistent Harmonic Approximation~\cite{Gillis1968,Werthamer1970,Bianco2017,Tadano2018,Monacelli2018} and defines a free energy upper bound (see Eq.~\eqref{eq:ineqGB}), the counterpart of which is known as ELBO (Evidence Lower Bound) in the statistics literature on variational inference. Conversely, minimizing the Gibbs--Bogoliubov free energy $\FeGB$ provides the best approximation of both the free energy $\Fe$ and the distribution $p(\Rvec)$.

We are searching for the $\gammavec$ parameters of the surrogate potential that minimize the KLD. The gradient of the KLD with respect to the parameters $\gammavec$ is:
\begin{equation}
\begin{split}
\nabla_{\gammavec} \DKL =& -\beta \nabla_{\gammavec} (\Fe - \FeGB) \\
                        =& \beta \nabla_{\gammavec} \FeGB \\
                        =& \beta \nabla_{\gammavec} \FeGB^0 + \beta \nabla_{\gammavec} \bigg[\braket{V(\Rvec) - \Vmlip(\Rvec)}_{\Vmlip}\bigg] \\
                        =& \beta \nabla_{\gammavec} \bigg(-\kBT\ln \bigg[ \int \dint \Rvec \mathrm{e}^{-\beta \Vmlip(\Rvec)} \bigg]\bigg) \\
                        &+\beta \nabla_{\gammavec} \bigg[\frac{1}{\widetilde{\Zpart}_{\gammavec}} \int \dint \Rvec V(\Rvec) \mathrm{e}^{-\beta \Vmlip(\Rvec)}
                        - \frac{1}{\widetilde{\Zpart}_{\gammavec}} \int \dint \Rvec \Vmlip(\Rvec) \mathrm{e}^{-\beta \Vmlip(\Rvec)}\bigg] \\
                        =&\beta \braket{\Xmlip(\Rvec)^\mathrm{T}}_{\Vmlip} +\beta \bigg[ \beta \bigg( \braket{\Xmlip(\Rvec)^\mathrm{T}}_{\Vmlip} \braket{ V(\Rvec)}_{\Vmlip} - \braket{ V(\Rvec)\Xmlip(\Rvec)^\mathrm{T}}_{\Vmlip} \bigg) \\
                        &- \beta \bigg( \braket{\Xmlip(\Rvec)^\mathrm{T}}_{\Vmlip} \braket{ \Vmlip(\Rvec)}_{\Vmlip} - \braket{ \Vmlip(\Rvec)\Xmlip(\Rvec)^\mathrm{T}}_{\Vmlip} \bigg) - \braket{\Xmlip(\Rvec)^\mathrm{T}}_{\Vmlip} \bigg] \nonumber \\
                        =&\beta^2 \bigg[ \braket{\Xmlip(\Rvec)^\mathrm{T}}_{\Vmlip} \big( \braket{ V(\Rvec)}_{\Vmlip} - \braket{ \Vmlip(\Rvec)}_{\Vmlip}\big) \\
                        &- \braket{ V(\Rvec)\Xmlip(\Rvec)^\mathrm{T}}_{\Vmlip} + \braket{ \Vmlip(\Rvec)\Xmlip(\Rvec)^\mathrm{T}}_{\Vmlip} \bigg] \mbox{\; .} \nonumber \\
\end{split}
\end{equation}
Since $\DKL$ and $\nabla_{\gammavec} \DKL$ are invariant with respect to the addition of a constant to the surrogate potential $\Vmlip$, one can impose without loss of generality that $\braket{ V(\Rvec)}_{\Vmlip}=\braket{\Vmlip(\Rvec)}_{\Vmlip}$ for a unique thermodynamic point (\emph{i.e.}~$\beta$ fixed), and obtain :
\begin{equation}
        \nabla_{\gammavec} \DKL = -\beta^2 \bigg[ \Braket{ \Xmlip(\Rvec)^\mathrm{T}V(\Rvec)}_{\Vmlip}  - \Braket{\Xmlip(\Rvec)^\mathrm{T}\Vmlip(\Rvec)}_{\Vmlip} \bigg] \mbox{\; ,} \nonumber
\end{equation}
with $\Vmlip(\Rvec)=\Xmlip(\Rvec)\gammavec$ the expression of the surrogate potential. Therefore, minimizing the Kullback--Leibler divergence ($\nabla_{\gammavec} \DKL =0$), which is equivalent to minimize the Gibbs--Bogoliubov free energy ($\nabla_{\gammavec} \FeGB = 0$), leads to the following optimal parameters:
\begin{equation}
\label{eq:LS_KB}
        \gammavec = \Braket{\Xmlip(\Rvec)^\mathrm{T} \Xmlip(\Rvec)}_{\Vmlip}^{-1}
                   \Braket{ \Xmlip(\Rvec)^\mathrm{T}  V(\Rvec)}_{\Vmlip}  \mbox{\; .}
\end{equation}
Note that we assumed implicitly that the matrix~$\Braket{\Xmlip(\Rvec)^\mathrm{T} \Xmlip(\Rvec)}_{\Vmlip}$ is invertible. We observed in all our simulations that this was always the case; in case it would not be, one would need to add a small regularization.

The non-trivial least-squares solution~\eqref{eq:LS_KB} shows a circular dependency over $\gammavec$ (the parameters $\gammavec$ appear in the right hand side of Eq.~\eqref{eq:LS_KB} through the effective potential used to perform the average) and can be solved using a self-consistent procedure. Thus, the initial problem turns into an optimisation problem, solved with a flexible and simple MLIP potential, using variational inference~\cite{Blei_2017,Yang_ArXiv_2019}. 

So far, only the DFT data $V(\Rvec)$ are used, and not their derivatives. In order to lower the number of DFT calculations needed, we also aim to utilize the forces $\Fvec=-\nabla_{\Rvec}V(\Rvec,\epsilonvec)$ and stresses $\pmb\sigma=\nabla_{\epsilonvec}V(\Rvec,\epsilonvec)/\Omega(\epsilonvec)$ of the target system (with $\epsilonvec$ giving the cell strains and $\Omega(\epsilonvec)$ the cell volume) to determine the surrogate equilibrium distribution and thereby obtain the $\gammavec$ parameters that enable the reproduction of these quantities. 
Let us consider the framework of the isothermal-isobaric (NPT) ensemble: the geometry of the periodic supercell is defined by a $3\times3$ matrix $\mathbf{h}$ and can change under the action of the internal or external pressure $p$. We then parametrize such changes as $\mathbf{h(\epsilonvec)}=\mathbf{h_0}(\mathbb{I}+\epsilonvec)$ with $\mathbf{h_0}$ the $3\times3$ equilibrium cell matrix, so that $\Omega (\epsilonvec)=\mathrm{det}[\mathbf{h(\epsilonvec)}]$ defines the volume (see Ref.~\cite{Cajahuaringa_JCP149,Kobayashi_JCP155_2021}). 
In this context, the target and surrogate distributions become $p(\Rvec, \epsilonvec)=\mathrm{e}^{-\beta (V(\Rvec)+p \Omega(\epsilonvec))}/\Zpart^{'}$ and $\qmlip(\Rvec, \epsilonvec) =\mathrm{e}^{-\beta (\Vmlip(\Rvec)+p \Omega(\epsilonvec))}/\widetilde{\Zpart}^{'}_{\gammavec}$, with the following target and surrogate configurational partition functions $\Zpart^{'}= \int\dint\epsilonvec\Zpart(\epsilonvec)\mathrm{e}^{-\beta p \Omega(\epsilonvec)}$ and $\widetilde{\Zpart}^{'}_{\gammavec}= \int\dint\epsilonvec\widetilde{\Zpart}_{\gammavec}(\epsilonvec)\mathrm{e}^{-\beta p \Omega(\epsilonvec)}$, respectively. 

The aim is to obtain an equation equivalent to Eq.~\eqref{eq:LS_KB} using information on the gradients. If the Kullback--Leibler divergence $\DKL$ gauges the discrepancy between the two distributions $\qmlip$ and $p$, the Fisher divergence $\DF^{\etavec}$~\cite{Yang_ArXiv_2019,Lyu_Proc_2009} measures the difference between these two distributions while including the gradient of the energy with respect to $\etavec$ (indexing the atomic positions $\Rvec$ or the cell strains $\epsilonvec$):
\begin{equation}
    \DF^{\etavec}(\qmlip \Vert p) = \iint\dint\Rvec\dint\epsilonvec \qmlip(\Rvec, \epsilonvec) \bigg\vert \nabla_{\etavec} \ln \bigg[ \frac{\qmlip(\Rvec, \epsilonvec)}{p(\Rvec, \epsilonvec)} \bigg] \bigg\vert^2 \geq 0 \mbox{\; .} \nonumber
\end{equation} 
To simplify the notation, we remove the arguments $(\Rvec, \epsilonvec)$ of the distributions, potentials, partition functions... in the following. This divergence has properties similar to the Kullback--Leibler one: the smaller $\DF^{\etavec}$ is, the closer the distribution $p$ and $\qmlip$ are, with $\DF^{\etavec}(\qmlip\Vert p)=0$ meaning that the two distributions are identical. This quantity can be rewritten as:
\begin{eqnarray}
    &\DF^{\etavec}(\qmlip \Vert p) = \displaystyle\iint\dint\Rvec\dint\epsilonvec \qmlip \bigg\vert \nabla_{\etavec} \bigg[\ln \bigg(\displaystyle \frac{\Zpart^{'}}{\widetilde{\Zpart^{'}_{\gammavec}}}\bigg)-\beta \Vmlip +\beta V\bigg]  \bigg\vert^2 \nonumber \\
    &= \beta^2 \displaystyle\iint\dint\Rvec\dint\epsilonvec \qmlip \bigg\vert \nabla_{\etavec}V - \nabla_{\etavec}\Vmlip \bigg \vert^2 \mbox{\; .}
\label{eq:DF_eta}
\end{eqnarray}
Depending on whether we consider the gradients with respect to the atomic positions $\Rvec$ or to the cell strains $\epsilonvec$, we obtain the following equations:
\begin{eqnarray}
    \DF^{\Fvec}(\qmlip \Vert p)    & = & \displaystyle\beta^2\braket{ \vert \Fvec     - \widetilde{\Fvec}_{\gammavec} \vert^2 }_{\Vmlip}  \mbox{\; ,} \\ 
    \DF^{\sigmavec}(\qmlip \Vert p) & = & \displaystyle\beta^2\braket{ \vert \sigmavec - \widetilde{\sigmavec}_{\gammavec} \vert^2 }_{\Vmlip} \mbox{\; ,}
\end{eqnarray}
with $\widetilde{\Fvec}_{\pmb{\gamma}}=-\nabla_{\Rvec}\Vmlip(\Rvec,\epsilonvec)$ the surrogate atomic forces and $\widetilde{\pmb\sigma}_{\pmb{\gamma}}=\nabla_{\epsilonvec}\Vmlip(\Rvec,\epsilonvec)/\Omega(\epsilonvec)$ the surrogate cell stresses, respectively.

As for the Kullback--Leibler divergence, we can compute the gradient of $\DF^{\etavec}$ to find out the $\gammavec$ parameters, solutions of the minimization procedure. 
Starting from Eq.~\eqref{eq:DF_eta}, and use once again $\nabla_{\gammavec} \Vmlip = \Xmlip^{\mathrm{T}}$, we obtain
\begin{eqnarray}
    \nabla_{\gammavec} \DF^{\etavec} & = & \nabla_{\gammavec} \bigg(\beta^2\braket{(\nabla_{\etavec}\Vmlip - \nabla_{\etavec}V)(\nabla_{\etavec}\Vmlip - \nabla_{\etavec}V)}_{\Vmlip} \bigg) \nonumber \\
    & = & \beta^2 \bigg( 2\braket{(\nabla_{\etavec}\Vmlip - \nabla_{\etavec}V)\nabla_{\etavec}(\nabla_{\gammavec}\Vmlip)}_{\Vmlip} \nonumber \\
    &   & -\beta \braket{\vert\nabla_{\etavec}\Vmlip - \nabla_{\etavec}V\vert^2 (\Xmlip^\mathrm{T}-\braket{\Xmlip^\mathrm{T}}_{\Vmlip})}_{\Vmlip}  \bigg) \mbox{\; .} 
\end{eqnarray}
The second term on the right-hand side of the previous equation comes from the gradient of the Boltzmann weight and is a lower-order term, as it is quadratic in $(\nabla_{\etavec}\Vmlip - \nabla_{\etavec}V)$. This term is much lower than the first one if the convergence towards the fixed point is reached~\cite{Yang_ArXiv_2019}, so it is discarded in the present version of \textsc{Mlacs}. If we impose that $\nabla_{\gammavec} \DF^{\etavec}=0$, we finally state that the $\gammavec$ parameters have to fulfill a non-trivial least-squares solution
\begin{equation}
\label{eq:LS_Fisher}
    \gammavec = \Braket{\nabla_{\etavec}\Xmlip^\mathrm{T} \nabla_{\etavec}\Xmlip}_{\Vmlip}^{-1} \Braket{ \nabla_{\etavec}\Xmlip^\mathrm{T} \nabla_{\etavec}V}_{\Vmlip} \mbox{\; ,}  
\end{equation}
equivalent to Eq.~\eqref{eq:LS_KB}, but using the forces or stresses $\nabla_{\etavec}V$ coming from DFT calculations. 

\subsection{Implementation}

In this section, we discuss how to solve Eq.~\eqref{eq:LS_KB} and~\eqref{eq:LS_Fisher}. While these formulas are similar to the ones obtained with a simple linear least-squares, they show a circular dependency which requires a self-consistent procedure. The aim of this section is also to provide a comprehensive formulation of the least-squares solution coming from the minimization of the following cost function: 
\begin{equation}
\Delta = \alpha_{\mathrm{E}} \DKL + \alpha_{\mathrm{F}} \DF^{\Fvec} + \alpha_{\mathrm{S}} \DF^{\sigmavec} \mbox{\; ,} 
\label{eq:costfunc}
\end{equation}
with $\alpha_{\mathrm{E}}$, $\alpha_{\mathrm{F}}$ and $\alpha_{\mathrm{S}}$ parameters allowing to adjust the contributions from the energies, forces and stresses. We next consider that we no longer have an infinity of atomic configurations but $N$ configurations distributed according to $w_n$ (see Eq.~\eqref{eq:average}), with $\Rvec_n$ and $\epsilonvec_n$ the atomic positions and strains of the $n^{th}$ configuration. In this context, $V_n = V(\Rvec_n,\epsilonvec_n)$, $\Fvec_n = -\nabla_{\Rvec}V_n$ (a vector with $3\Nat$ components) and $\sigmavec_n = \nabla_{\epsilonvec}V_n/\Omega(\epsilonvec_n)$ (a vector with 6 components in the Voigt notation), are respectively the energy, the forces and the stresses of the $n^{th}$ configuration. Thus, we can construct the $(3\Nat + 7)N$-long label vector $\mathbf{Y}_N$ with energies, forces and stresses of the $N$ configurations:
\begin{equation}
\mathbf{Y}_N = 
    \begin{bmatrix}
        V_1 , \Fvec_1 , \sigmavec_1 , \ldots , V_{N} , \Fvec_{N} , \sigmavec_{N}
    \end{bmatrix}^{\mathrm{T}} \mbox{\; .}
\end{equation}
For the MLIP potential, we define $\Xd_{n,k} = \Xd_k(\Rvec_n,\epsilonvec_n)$, $\mathbf{f}_{n,k}= -\nabla_\Rvec\Xd_{n,k}$ and $\mathbf{s}_{n,k}= \nabla_{\epsilonvec} \Xd_{n,k}/\Omega(\epsilonvec_n)$, respectively the $k^{th}$ descriptor, its gradient with respect to the atomic positions and with respect to the cell strains for the $n^{th}$ configuration. We can then write the $(3\Nat + 7) N\times K$ feature matrix $\mathbf{X}_N$ built using the descriptors:
\begin{equation}
{\mathbf{X}_N} = 
    \begin{bmatrix}
        \Xd_{1,1}        & \hdots & \Xd_{1,K} \\
        \mathbf{f}_{1,1} & \hdots & \mathbf{f}_{1,K} \\
        \mathbf{s}_{1,1} & \hdots & \mathbf{s}_{1,K} \\
        \vdots & \vdots & \vdots \\
        \vdots & \vdots & \vdots \\
        \Xd_{N,1}        & \hdots & \Xd_{N,K} \\
        \mathbf{f}_{N,1} & \hdots & \mathbf{f}_{N,K} \\
        \mathbf{s}_{N,1} & \hdots & \mathbf{s}_{N,K} \\
    \end{bmatrix} \mbox{\; ,}
\end{equation}
and the $(3\Nat + 7) N\times (3\Nat + 7) N$ diagonal matrix $\mathbf{W}_N$ including the regularization parameters (the weights) as:
\begin{equation}
\mathbf{W}_N=\mathrm{diag}\bigg\{\alpha_{\mathrm{E}}w_1,\alphavec_{\mathrm{F}}w_1,\alphavec_{\mathrm{S}}w_1,\hdots, \alpha_{\mathrm{E}}w_N,\alphavec_{\mathrm{F}}w_N,\alphavec_{\mathrm{S}}w_N\bigg\} \label{eq:Weights}
\end{equation}
with $\alphavec_{\mathrm{F}}=\big[\alpha_{\mathrm{F}},...,\alpha_{\mathrm{F}}\big]$ a vector with $3\Nat$ components and $\alphavec_{\mathrm{S}}=\big[\alpha_{\mathrm{S}},...,\alpha_{\mathrm{S}}\big]$ a vector with 6 components. Using these definitions, the $\widehat{\gammavec}_N$ parameters, solutions to the least-squares problem associated with Eq.~\eqref{eq:costfunc}, are given by
\begin{equation}
\label{eq:WLSN}
    \widehat{\gammavec}_N = (\mathbf{X}_N^{\mathrm{T}}\mathbf{W}_N\mathbf{X}_N)^{-1}(\mathbf{X}_N^{\mathrm{T}}\mathbf{W}_N\mathbf{Y}_N) \mbox{\; ,}
\end{equation}
with $\widehat{\gammavec}_N = \begin{bmatrix} \widehat{\gamma}_1 , \widehat{\gamma}_2 , \ldots , \widehat{\gamma}_{k} \end{bmatrix}^{\mathrm{T}}$ and $\mathbf{X}_N^{\mathrm{T}}\mathbf{W}_N\mathbf{Y}_N$ column vectors with $K$ components, and $\mathbf{X}_N^{\mathrm{T}}\mathbf{W}_N\mathbf{X}_N$ a $K\times K$ matrix. 

\begin{figure}[h]
\includegraphics[width=\textwidth]{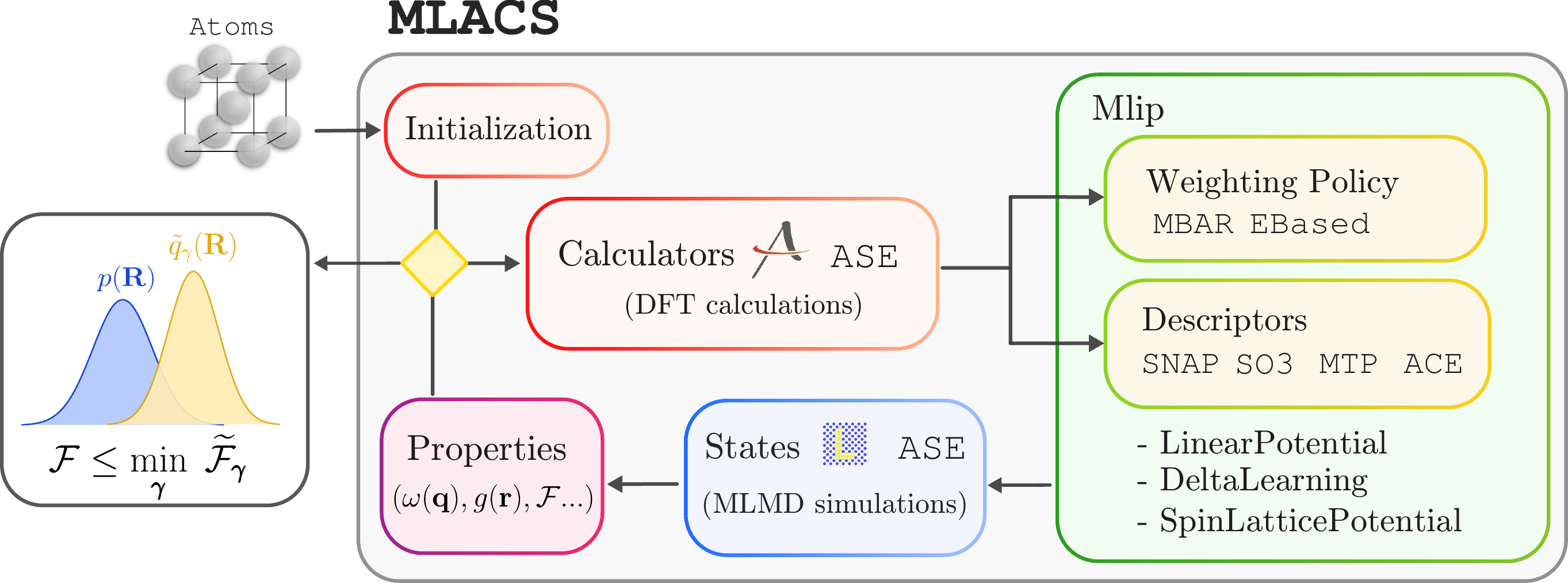}
\caption{\label{fig:MLACS} Workflow of \textsc{Mlacs}.}
\end{figure}
Since the right-hand side of Eq.~\eqref{eq:WLSN} implicitly depends on the $\widehat{\gammavec}_N$ parameters (the reweighted configurations being distributed according to $\ensuremath{\widetilde{q}_{\widehat{\gammavec}_N}}(\Rvec)$), a self-consistent (SC) procedure is employed. The algorithm implemented in our python code is detailed in Alg.~\ref{alg:mlacs} (see also Fig.~\ref{fig:MLACS}). 
Basically, the \textsc{Mlacs} SC loop proceeds using an active learning strategy as follows: $(i)$ a first atomic configuration corresponding to atomic positions randomly generated around equilibrium is produced (\textsc{Mlacs} is also able to start using a previous MLIP), $(ii)$ DFT and MLIP single point calculations are performed using this atomic configuration, $(iii)$ the weight of this configuration is computed (see below for more information), $(iv)$ physical quantities can be obtained using the weights and a criterion based on the desired property, for example phonon frequencies ($\frac{\Delta\omega}{\omega} \leq 1\%$) for solid or pair distribution function ($\Delta g(r)\leq 0.1$) for liquid, is used to stop the SC loop, $(v)$ the coefficients of the MLIP are computed, $(vi)$ a MD simulation using the MLIP (MLMD) is then performed during 0.1-1 ps with \textsc{Lammps}~\cite{Thompson2022} (Large-scale Atomic/Molecular Massively Parallel Simulator), a classical molecular dynamics program that solves Newton's equations of motion, $(vii)$ the last atomic configuration is extracted, and finally the calculation restarts at the beginning of the loop. 
\%
\begin{algorithm}[H]
\caption{\textsc{Mlacs}}\label{alg:mlacs}
\begin{algorithmic}[1]
\REQUIRE Start with a MLIP $\widehat{\gammavec}_1$ or with atomic configurations $\{\Rvec_1,\epsilonvec_1\}$ randomly distributed around equilibrium positions and cell shape.\\ 
\underline{Loop over SC steps}
\FOR{$N=1,2\ldots$}
\STATE \underline{DFT}: $V_N = V(\Rvec_N,\epsilonvec_N$), $\Fvec_N = -\nabla_{\Rvec}V_N$ and $\sigmavec_N= \nabla_{\epsilonvec}V_N$
\STATE \underline{MLIP}: $\Xd_N= \{\Xd_k(\Rvec_N,\epsilonvec_N)\}_{k=1, ...K}$, $\mathbf{f}_N= -\nabla_\Rvec\Xd_N$ and $\mathbf{s}_N= \nabla_{\epsilonvec} \Xd_N$ \\
\FOR{$n=1,\ldots$, $N$}
\STATE \underline{Reweighting (MBAR)}: compute $\widetilde{V}_{\widehat{\gammavec}_N}(\Rvec_{n},\epsilonvec_{n})$, $\widetilde{q}_{\widehat{\gammavec}_N}(\Rvec_{n},\epsilonvec_{n})$ and $w_n^N$ 
\ENDFOR 
\STATE Compute $\braket{O}_N = \sum_{n} O(\Rvec_n,\epsilonvec_n) w_n^N$ and $\Delta\braket{O}=\vert \braket{O}_N-\braket{O}_{N-1} \vert$
\STATE {\bf if} ($\Delta\braket{O}\leq$ criterion) {\bf exit}
\STATE Build $\mathbf{W}_N=\mathrm{diag}\big\{\alpha_{\mathrm{E}}w_1^N,\alphavec_{\mathrm{F}}w_1^N,\alphavec_{\mathrm{S}}w_1^N,\hdots, \alpha_{\mathrm{E}}w_N^N,\alphavec_{\mathrm{F}}w_N^N,\alphavec_{\mathrm{S}}w_N^N\big\}$
\STATE Build $\Xvec_N=\{\Xvec_{N-1},\Xd_{N},\mathbf{f}_{N},\mathbf{s}_{N}\}$ and $\Yvec_N=\{\Yvec_{N-1},V_{N},\Fvec_{N},\sigmavec_{N}\}$\\
\STATE \underline{MLIP fit}: Compute $\widehat{\gammavec}_N = (\Xvec_N^{\rm T}\Wvec_N\Xvec_N)^{-1} (\Xvec_N^{\rm T}\Wvec_N\Yvec_N)$
\STATE \underline{MLMD simulation}: using $\widetilde{V}_{\widehat{\gammavec}_N}(\Rvec,\epsilonvec)=\sum_{k=1}^{K} \widehat{\gamma}_{N,k} \Xd_k(\Rvec,\epsilonvec)$ 
\STATE Extract the last configuration $\{\Rvec_{N+1},\epsilonvec_{N+1}\}$ from MLMD trajectory
\ENDFOR
\end{algorithmic}
\end{algorithm}

A crucial reweighting procedure is used within the algorithm. Since the potential is evolving at every step of the SC procedure, the configurations generated previously are not distributed according to the canonical distribution of the current potential. In order to correctly keep these configurations in the database and represent the configuration space according to the current potential, a reweighting strategy has been implemented, based on the the Multistate Bennett Acceptance Ratio (MBAR)~\cite{Shirts2008,shirts2017reweighting,Stoltz_2010}. In this framework, the canonical equilibrium distribution is computed for every configuration $n$ according to the potential at every step $N$. The weight $w_n^N$ of the configuration $n$ according to the potential built at the step $N$ is
\begin{equation}
\label{eq:MBAR}
w_n^N=\frac{\displaystyle{\widetilde{q}_{\widehat{\gammavec}_N}\big(\Rvec_{n},\epsilonvec_{n}\big)}}{\displaystyle{\sum_{k=0}^{N}\widetilde{q}_{\widehat{\gammavec}_k}\big(\Rvec_{n},\epsilonvec_{n}\big)}} \mbox{\; ,}
\end{equation}
with the following non-linear equation defining the partition function 
\begin{equation}
\ensuremath{\mathcal{\widetilde{Z}}_{\widehat{\gammavec}_N}} = \sum_{n=0}^N \frac{\displaystyle{e^{-\beta \big(\widetilde{V}_{\widehat{\gammavec}_N}(\Rvec_{n}) +p\Omega(\epsilonvec_{n})\big)}}}{\displaystyle{\sum_{k=0}^N e^{-\beta \big(\widetilde{V}_{\widehat{\gammavec}_k}(\Rvec_{n}) +p\Omega(\epsilonvec_{n})\big)}}/ \ensuremath{\mathcal{\widetilde{Z}}_{\widehat{\gammavec}_k}}}
\end{equation}
associated with the surrogate distribution $\widetilde{q}_{\widehat{\gammavec}_N}\big(\Rvec_{n},\epsilonvec_{n}\big) = \displaystyle{{e^{-\beta \big(\widetilde{V}_{\widehat{\gammavec}_N}(\Rvec_{n})+p\Omega(\epsilonvec_{n})\big)}}/\ensuremath{\mathcal{\widetilde{Z}}_{\widehat{\gammavec}_N}}}$. 
The matrix $\Wvec_N$ defined by Eq.~\eqref{eq:Weights} becomes:
\begin{equation}
	\mathbf{W}_N=\mathrm{diag}\bigg\{\alpha_{\mathrm{E}} w_1^N,\alphavec_{\mathrm{F}} w_1^N,\alphavec_{\mathrm{S}} w_1^N,\hdots, \alpha_{\mathrm{E}} w_N^N,\alphavec_{\mathrm{F}} w_N^N,\alphavec_{\mathrm{S}} w_N^N\bigg\} \mbox{\; .}
\end{equation}
Using this definition of weights, we are able to reuse the data computed for all the SC steps. These weights are employed to fit the potential for the next SC step and to compute physical properties. This strategy strongly reduces the number of configurations needed to achieve the convergence.

MBAR weights constrain the training set to configurations representative of the target ensemble. It reduces the configuration space on which the potential is trained, giving insight into why relatively simple potentials suffice for most applications in \textsc{Mlacs}. Reweighting plays a critical role in active learning strategies, where initial configurations may be irrelevant to the target ensemble and could otherwise complicate the fitting procedure. 

\section{Free energy computation}
\label{sec:free energy computation}

\subsection{Theory}

Free energy is one of the most important quantity in materials science. However, it is not possible to obtain this quantity using the standard numerical methods of statistical physics, such as Molecular Dynamics or Monte Carlo, because it cannot be expressed as the statistical average of some microscopic quantity. An analytical formulation is available for harmonic systems, but in the general case, there is no analytical formulation fully accounting for anharmonic effects. One possibility is to compute this quantity using thermodynamic integration but this kind of calculation requires so far tremendous computational resources for achieving {\it ab initio} accuracy~\cite{Alfe1999, Vocadlo2002, Soubiran2020, Alfe2022, Jung2023}. With \textsc{Mlacs}, computing the free energy of a system (solid or liquid) through TI is made straightforward. Utilizing an optimal and local (for a given crystallographic phase and thermodynamic point) MLIP potential $\Vmlip$ that closely reproduces the target equilibrium distribution makes this computation feasible while maintaining {\it ab initio} accuracy. 
Moreover, this strategy partially cancels both finite-time and finite-size effects, which are often an issue in free energy calculations, making this computation feasible for large systems. 
For example, computing the free energy of gold in the bcc phase (8192 atoms in the supercell) is completed in less than 10 hours on 128 processors.

TI simulations enable to calculate the free energy difference between the ``system of interest'' (\textit{ab initio} system) and a ``reference system'' (for which the free energy can be exactly calculated) by connecting these two systems along a non-physical path which slowly changes the description from one to the other. Technically, one can build a parameterized Hamiltonian $H(\lambda)$, linking the Hamiltonians of the reference $H_{\text{ref}}$ and interest $H_{\text{int}}$ systems with a coupling parameter $\lambda$ varying from 0 to 1, such as 
\begin{equation}
    H(\lambda) = \lambda H_{\text{int}} + (1 - \lambda) H_{\text{ref}}  \text{ .}
\end{equation}
For solids, the Frenkel-Ladd path~\cite{Frenkel1984, Frenkel, Koning1996, Vega2007} uses the Einstein crystal as the reference system. The (classical) free energy $\Fe_{\text{ref}}^{\text{Ein}}(\Omega,T)$ of such a system is analytically known as
\begin{equation}
    \Fe_{\text{ref}}^{\text{Ein}}=3N_{\text{at}}k_{B}T \ln \left( \frac{\hbar \omega_E}{k_{B} T}\right ) \text{,}
    \label{eq:einstein}
\end{equation}
with $\omega_E=\sqrt{k_E / m}$ the frequency of the Einstein modes. In theory, any value of the spring constant $k_E$ could be used, but it is more convenient, in order to have an as smooth as possible path connecting the reference and interest systems, to fix the spring constant to a value that best mimics the vibrations of the interest system. Hence, assuming that the harmonic approximation reasonably holds, the spring constant $k_E$ is computed using the equipartition theorem $\frac{3}{2}\mathrm{k_B T} = \frac{1}{2} k_E \langle (\Delta \Rvec )^2\rangle$ \cite{Frenkel}, and therefore reflects vibration frequencies through the mean squared displacement of atoms $\langle (\Delta \Rvec )^2\rangle$.

If we consider a liquid system, the choice of a reference system is less obvious. While the free energy of the ideal gas is analytically known, it is not advisable to consider it as a reference system because a liquid-vapor transition could occur during TI~\cite{Abramo2015}.
Several methods are proposed to alleviate this issue, such as adding an intermediate stage in the switching process passing through a purely repulsive system before the ideal gas~\cite{Broughton1983, Broughton1987}, by optimizing an empirical reference potential~\cite{Zhu2017}, by using inverse powered pair potentials~\cite{Alfe1999, Alfe2000, Alfe2002, Alfe2002_2, Vocadlo2002} or Lennard--Jones fluid~\cite{Wijs1998} for which the free energy is known~\cite{Johnson1993}. \textsc{Mlacs} uses the Uhlenbeck--Ford (UF) model as reference system~\cite{Allen, UFM, Leite2016, Leite2017}, which is basically a purely repulsive potential for interacting particles of fluids (a soft-sphere)
\begin{equation}
    U_{\text{UF}}(r) = -\frac{p}{\beta}\ln \left( 1 - \exp{\left ( -\frac{r}{\sigma} \right )^2} \right ) \text{,}
\end{equation}
with $r$ the distance between atoms, $\sigma$ the length-scale and $p$ a positive scaling factor that determines the strength of interactions. It decays rapidly for increasing interatomic distances and presents a logarithmic divergence near the origin. Two main reasons make this model an ideal reference system. The first one is because it is always a fluid (for $p \leq 100$ ~\cite{Leite2017, Leite_2019}) avoiding liquid-vapor transition during the switching process of TI. The second is that the form of the potential makes excess free energy expressible as a function of the thermodynamic dimensionless density $\rho$ parameter as
\begin{equation}
    \Fe^\text{UF}_{\text{ref}} (\sigma, p, T) = k_B T \sum_{n=1}^{\infty}\frac{B_{n+1}(p)}{n} \left ( \frac{(\pi \sigma^2)^{3/2}}{2}\rho \right )^n \text{,}
\end{equation}
where $\frac{(\pi \sigma^2)^{3/2}}{2}$ and $B_{n+1}(p)$ are the reduced virial coefficients, which can be exactly evaluated~\cite{Leite2016, Leite_2019}. Tabulated UF potentials are available in \textsc{Mlacs} for $p= 1, 25, 50, 75, 100$ taken from Leite \textit{et al.}~\cite{Leite2016}. 

The TI method computes the free energy difference corresponding to the alchemical transformation between two thermodynamic states described by $\lambda=0$ and $\lambda=1$ as
\begin{align}
    \Delta \Fe_{\text{ref} \rightarrow \text{int}} & = \Fe_{\text{int}} - \Fe_{\text{ref}} \text{ ,}\\
    & = \Fe(\lambda=1) - \Fe(\lambda=0) \text{ ,}\\
    & = \int_0^1 \frac{\partial \Fe(\lambda)}{\partial \lambda} d\lambda \text{ .}
    \label{eq:delta_f}
\end{align}
Knowing that $\Fe(\lambda) = -\mathrm{k_B T} \ln Z(\lambda)$, where $Z(\lambda)= \frac{1}{h^{3N}}\int_{\mathbf{R}, \mathbf{P}} \text{e}^{-\beta H(\mathbf{R}, \mathbf{P}; \lambda)} d\mathbf{R} d\mathbf{P}$ is the canonical partition function, one obtains 
\begin{equation}
   \Delta \Fe_{\text{ref} \rightarrow \text{int}} = \int_0^1 \left \langle \frac{\partial H(\lambda)}{\partial \lambda} \right \rangle _{\lambda} d\lambda \text{ ,}
   \label{eq:delta_f_eq}
\end{equation}
with $\langle \cdots \rangle_{\lambda}$ the canonical statistical average for one value of $\lambda$ with the canonical measure $Z(\lambda)^{-1}\text{e}^{-\beta H(\mathbf{R}, \mathbf{P}; \lambda)}d\mathbf{R} d\mathbf{P}$. The reversible work  $W_\text{rev}$ associated with this process is equal to the free energy difference between the two equilibrium systems~\cite{Frenkel}: $W_{\text{rev}} \equiv \Delta \Fe_{\text{ref} \rightarrow \text{int}}$.
In practice, the transformation path is discretized into small intervals $\Delta\lambda$, and the smaller $\Delta\lambda$, the more accurate the free energy. This makes standard TI an expensive method because each generated system along a reversible path has to reach equilibrium. 

Nonequilibrium thermodynamic integration~\cite{Watanabe1990} (NETI) overcomes this issue by linking reference and interest systems along a single classical simulation in which the explicitly time-dependent coupling parameter $\lambda(t)$ goes from 0 to 1 during a switching time $t_s$. Only states at $\lambda=0$ are at equilibrium, making the whole trajectory a nonequilibrium process which deviates from the quasi-static path. If the transformation is very slow, the system remains close to equilibrium. Nevertheless, numerical simulation only allows a finite time $t_s$, making the transformation non reversible, with an associated irreversible work $W_{\text{irr}}$. Jarzynski's identity~\cite{Jarzynski1997, Jarzynski2} $\text{e}^{-\beta \Delta \Fe} = \overline{\text{e}^{-\beta W_{\text{irr}}}}$ establishes a theoretical framework that allows expressing the free energy difference between the initial and final states as an average $\overline{X}$ over various realizations of the irreducible work. In practice, the irreversible work 
\begin{equation}                                          
    W_{\text{irr}}=\int_{0}^{t_{s}} \frac{d \lambda(t)}{dt} \frac{\partial H(\mathbf{R}(t), \mathbf{P}(t);\lambda)}{\partial \lambda} \bigg|_{\lambda=\lambda(t)}  dt
    \label{eq:w_dyn}
\end{equation}
is evaluated using a parameter evolving continuously during the simulation and the reversible work becomes
\begin{equation}
    W_{\text{rev}} = \overline{W_{\text{irr}}} - \overline{E_d} \text{ ,}
\end{equation}
with $\overline{E_d}$ the dissipated energy depending on how fast the Hamiltonian is switched from one state to the other. This term is equal to zero only in the quasi-static limit ($t_{\text{s}} \rightarrow \infty$), 
but it has been proved~\cite{Koning2005} and shown~\cite{Freitas2016,Paula2019} that this error vanishes when executing the process in the forward and backward ways sufficiently slowly so that linear response theory is still valid ($E_{d} = E_{d}^{i \rightarrow f} = E_{d}^{f \rightarrow i}$).  Thus, by computing the average of $\Delta \Fe_{\text{ref} \rightarrow \text{int}}$ over both backward and forward directions,
\begin{equation}
    \Delta \Fe_{\text{ref} \rightarrow \text{int}} = \frac{1}{2} \left ( W_{\text{rev}}^{i \rightarrow f} - W_{\text{rev}}^{f \rightarrow i} \right ) = \frac{1}{2} \left ( W_{\text{irr}}^{i \rightarrow f} - W_{\text{irr}}^{f \rightarrow i} \right ) \text{ ,}
    \label{eq:wf}
\end{equation}
we can evaluate the free energy difference as the difference between irreversible works. Consequently by performing different independent runs, including each a forward and a backward process, we can compute the mean of this stochastic variable $W_{\text{irr}}$ and reduce the error due to the produced heat $E_d$. Finally, the free energy of the anharmonic surrogate system $\Fe_{\text{int}}$ (which is exacly the same quantity as $\FeGB^0$ seen above) can be estimated as the sum of the free energy of the reference system $\Fe_{\text{ref}}$ plus the free energy difference coming from the difference of the dynamic works:
\begin{equation}
    \Fe_{\text{int}} = \Fe_{\text{ref}} + \Delta \Fe_{\text{ref} \rightarrow \text{int}} \text{ .}
    \label{eq:tif}
\end{equation}
A correction $\delta \Fe$ due to the fixed center of mass constraint in MD \cite{Polson2000, Navascus2010} needs to be added in Eq.~\eqref{eq:tif}. The bigger the simulated box, the smaller this $\delta \Fe$.

The NETI calculations involving the surrogate potential are no longer time consuming since they are carried  without any {\it ab initio} calculations. There is still one step to perform in order to obtain properly the free energy $\Fe$ of the fully \textit{ab initio} system. The free energy difference $\Delta \Fe_{\text{int} \rightarrow \text{AI}}$ remaining between the MLIP and {\it ab initio} systems is evaluated using a cumulant expansion such as in free energy perturbation theory~\cite{Zwanzig1954}
\begin{equation}
   \Delta \Fe_{\text{int} \rightarrow \text{AI}}  = \Fe - \FeGB^0 = -\frac{1}{\beta} \ln \big\langle \mathrm{e}^{-\beta \Delta V(\Rvec)} \big\rangle_{\Vmlip} \mbox{\; ,}
\end{equation}
with $\Delta V(\Rvec)=V(\Rvec) - \Vmlip(\Rvec)$. This equation can be expanded in cumulants~\cite{Stoltz_2010} of the potential energy difference
\begin{equation}
\label{eq:cum exp}
   \Delta \Fe_{\text{int} \rightarrow \text{AI}} = \sum_{n=1}^{\infty} \frac{(-\beta)^{n-1} \kappa_n}{n!} \mbox{\; ,}
\end{equation}
where $\kappa_n$ is the $n$-th order cumulant of the potential energy differences.
Usually, the expansion~\eqref{eq:cum exp} is truncated at 2nd order, using only the first and second order cumulants of the potential energy difference: 
\begin{equation}
\begin{split}
    \kappa_1 &= \braket{\Delta V(\Rvec)}_{\Vmlip}  \mbox{\; ,}\\
    \kappa_2 &= \braket{\Delta V(\Rvec)^2}_{\Vmlip} - \braket{\Delta V(\Rvec)}_{\Vmlip}^2  \mbox{\; .}
\end{split}
\end{equation}
In this context, only an approximate free energy difference is obtained, except for systems where the potential energy difference follows a Gaussian distribution. Using this cumulant expansion, the free energy difference becomes 
\begin{equation}
    \Delta\Fe_{\text{int} \rightarrow \text{AI}} \approx \braket{\Delta V(\Rvec)}_{\Vmlip} -\frac{\beta}{2} \bigg(\braket{\Delta V(\Rvec)^2}_{\Vmlip} - \braket{\Delta V(\Rvec)}_{\Vmlip}^2\bigg)  \mbox{\; .} \label{eq:errorFe}
\end{equation}
Since the minimization of the Gibbs--Bogoliubov free energy is equivalent to the Kullback--Leibler Divergence minimization (see section~\ref{sec:mlacs}), one can relate these terms to the information in $p(\Rvec)$ not contained in $\widetilde{q}_{\gammavec}(\Rvec)$. Consequently, this second-order cumulant free energy difference gives not only a correction to the Gibbs--Bogoliubov free energy, but also measures the accuracy of \textsc{Mlacs}. The better the sampling done by \textsc{Mlacs}, the smaller $\Delta\Fe_{\text{int} \rightarrow \text{AI}}$. This quantity is zero if the {\it ab initio} $p(\Rvec)$ and MLIP $\qmlip(\Rvec)$ distributions are equal, and is generally small when the true and surrogate distributions are close to each other. In addition, since the only data needed to evaluate this correction are the energies coming from the atomic configurations already computed by the DFT and MLIP potentials, this quantity can be estimated during a \textsc{Mlacs} simulation at no extra cost.

\begin{figure}
    \centering
    \includegraphics[width=0.8\linewidth]{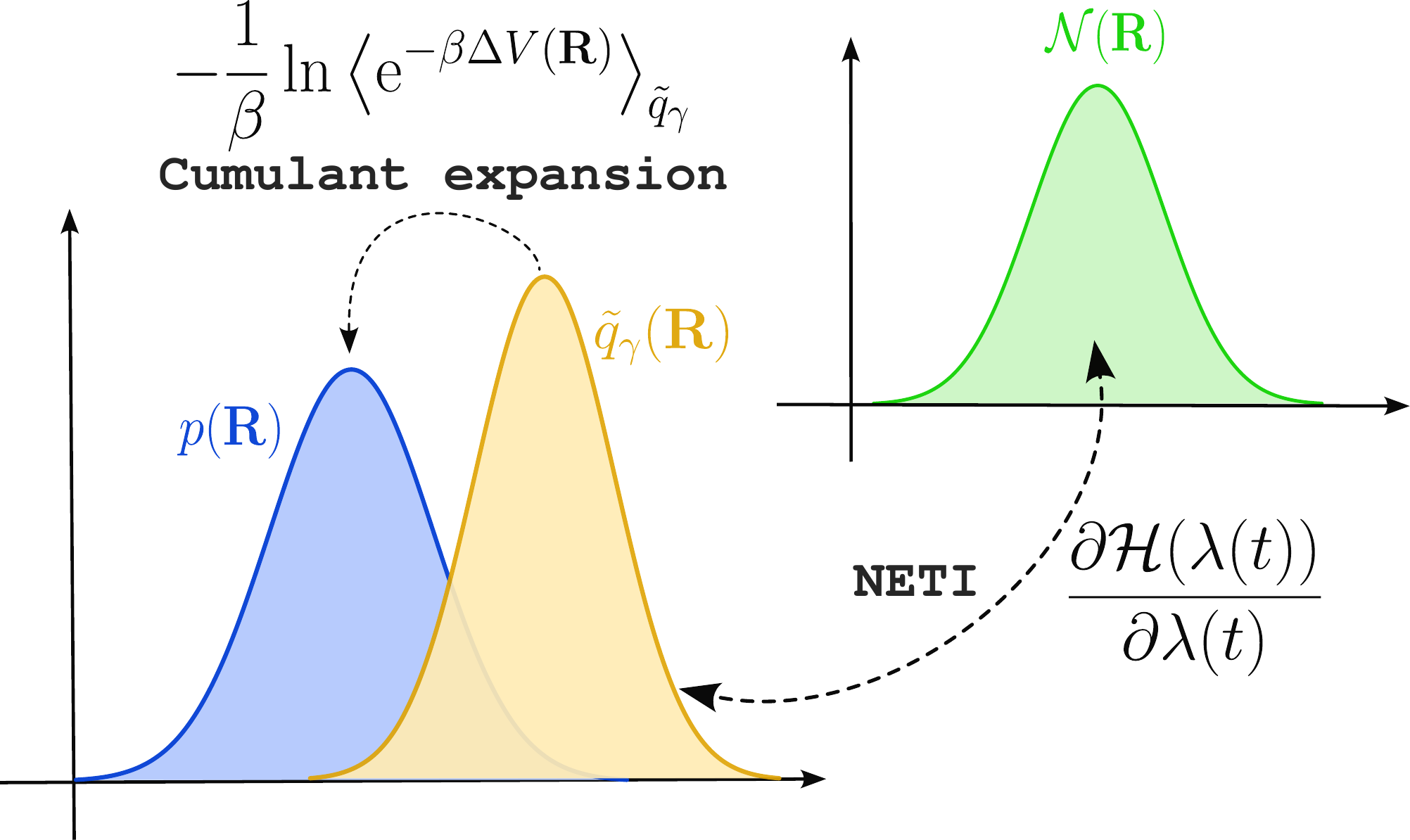}
    \caption{Evaluation of the {\it ab initio} free energy in two steps: first, using NETI simulations between the ``reference system'' (the Einstein or Uhlenbeck-Ford model in green) and the ``system of interest'' (the surrogate MLIP potential in yellow), and secondly, using a cumulant expansion between the MLIP ``system of interest'' and the {\it ab initio} ``target system'' (in blue).
    }
    \label{fig:timethod}
\end{figure}

Lastly, the free energy of the {\it ab initio} target system, as computed in this work, becomes
\begin{equation}
    \Fe = \Fe_{\text{ref}} + \Delta \Fe_{\text{ref} \rightarrow \text{int}} + \Delta\Fe_{\text{int} \rightarrow \text{AI}}
\end{equation}
and the above described procedure is illustrated in Fig.~\ref{fig:timethod}.

\subsection{Implementation}

\begin{figure}[h]
\includegraphics[width=\textwidth]{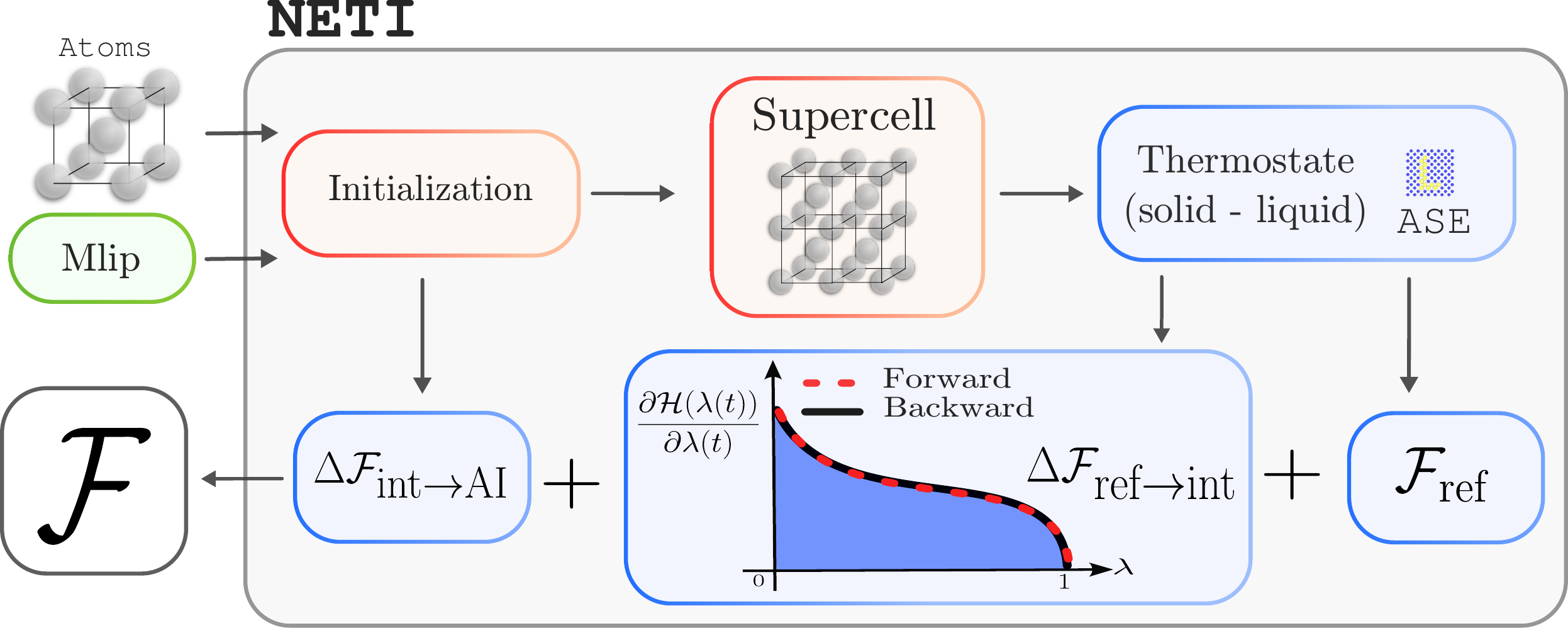}
\caption{\label{fig:NETI} Workflow of NETI in \textsc{Mlacs}.}
\end{figure}

The \textsc{Mlacs} package performs NETI simulations via an interface available in the \textsc{Lammps} code for both solid~\cite{Freitas2016} and liquid systems~\cite{Paula2019}. The \textsc{Lammps} input is automatically built to perform such a calculation by a properly defined object when calling the class dedicated to NETI in \textsc{Mlacs}. For more details about the NETI parameters (size effect, switching time...) to tune in \textsc{Lammps} for convergence, see Refs.~\cite{Freitas_2016, Leite_2019}. The workflow of the NETI implementation in \textsc{Mlacs} is illustrated in Fig.~\ref{fig:NETI}. We stress that the simulated structure can be built using the \textsc{Ase} environment~\cite{ase-paper} and that functions coming from the \textsc{Ase} object are used all along the process.

\begin{algorithm}[htp]
\caption{\textsc{NETI-\textsc{Mlacs} in the $N\Omega T$ canonical ensemble}}\label{alg:ti}
\begin{algorithmic}[1]
\REQUIRE $H_{\text{int}}=\frac{\mathbf{p}^2}{2 m_i} + \Vmlip(\Rvec)$ for a given thermodynamic state, an equilibrium structure $\{\Rvec_{eq}\}_{i=1,...N_{at}}$ and volume $\Omega_{eq}$ as initial state. \\ 
\underline{Initialization}
\IF{solid}
\STATE Compute $\langle (\Delta  \Rvec)^2\rangle = \langle (\Rvec - \Rvec_{eq})^2 \rangle$ and $k_E = \frac{3 k_B T}{\langle (\Delta \Rvec )^2\rangle}$
\STATE Set $H_{\text{ref}} \equiv \sum_{i=1}^{N_{at}} \left ( \frac{\mathbf{p}_i^2}{2 m_i} + \frac{1}{2}  m_{i} \omega_E^2 (\Delta\Rvec_i)^2\right )$
\ELSIF{liquid}
\STATE Compute the density $\rho$
\STATE Set $H_{\text{ref}} \equiv \sum_{i=1}^{N_{at}} \left ( \frac{\mathbf{p}_i^2}{2 m_i} -\frac{p}{\beta}\ln \left( 1 - \exp{\left ( -\frac{\Rvec_{i}}{\sigma} \right )^2} \right ) \right )$
\ENDIF 
\STATE Compute $\Fe_{\text{ref}}(\Omega, T)$\\
\underline{Loop over independent realizations}
\FOR{$j=1,2...$}
\STATE Equilibrate the initial state during $t_\text{eq}$. Set $W_{\text{irr}}^{i\rightarrow f} = 0$
\FOR{$k=1,2... {t_s\over\Delta t}$}
\STATE $W_{\text{irr}}^{i\rightarrow f} = W_{\text{irr}}^{i\rightarrow f} + \left( H_{\text{int}}(k\Delta t) - H_{\text{ref}}(k\Delta t)\right)$
\ENDFOR
\STATE Equilibrate the final state during $t_\text{eq}$. Set $W_{\text{irr}}^{f\rightarrow i} = 0$
\FOR{$k=1,2...{t_s\over\Delta t}$}
\STATE $W_{\text{irr}}^{f\rightarrow i} = W_{\text{irr}}^{f\rightarrow i} + \left( H_{\text{int}}(k\Delta t) - H_{\text{ref}}(k\Delta t)\right)$
\ENDFOR
\STATE Compute $\Delta \Fe_{\text{ref} \rightarrow \text{int}}^j =  \frac{1}{2} ( W_{\text{irr}}^{i \rightarrow f} - W_{\text{irr}}^{f \rightarrow i} )$ and $\overline{\Delta \Fe} = \frac{1}{j}\sum_{k=1}^{j} \Delta \Fe_{\text{ref} \rightarrow \text{int}}^i$
\STATE {\bf if} ($|\Delta \Fe_{\text{ref} \rightarrow \text{int}}^j - \overline{\Delta \Fe}| \leq$ criterion) {\bf exit}
\ENDFOR
\STATE $\Fe(\Omega, T) = \Fe_{\text{ref}}(\Omega, T) + \Delta \Fe_{\text{ref} \rightarrow \text{int}} + \Delta \Fe_{\text{int} \rightarrow \text{AI}}$
\STATE $G(p, T) = \Fe(\Omega, T) +  p\Omega_\text{eq}$ 
\end{algorithmic}
\end{algorithm}

For validation purposes, and to demonstrate the robustness of the NETI library in \textsc{Mlacs}, we present free energy results on both solid (iron) and liquid (water) systems. These data are compared to the ones obtained using the Python library \textsc{Calphy}~\cite{Menon2021}, which automatically performs NETI free energy calculations using the MD solvers available in the \textsc{Lammps} code. Iron is described using an EAM potential~\cite{Feeam} and water using a coarse-grained model. Results in table~\ref{tab:fe_results} show a good agreement between present calculations and previous studies for iron~\cite{Freitas_2016} and water~\cite{Leite_2019}, with meV precision on free energies. 

\begin{table*}[htp]
\centering
\caption{Free energy of iron (left side) and water (right-side) at $p=0$~GPa  for various temperatures, compared to values extracted from Ref.~\cite{Freitas_2016} and Ref.~\cite{Leite_2019}, respectively. 
}
\begin{tabular}{lccclccc} 
    \hline
    \hline
    & \multicolumn{3}{c}{$\Fe$ (eV/atom)} & & \multicolumn{3}{c}{$\Fe$ (eV/atom)} \\ 
    \cline{2-4} \cline {6-8}
    T(K) & \cite{Freitas_2016} & Present & $\Delta\times 10^{-3}$ & T(K) & \cite{Leite_2019} & Present & $\Delta\times 10^{-3}$ \\
    \hline
    400  & -4.3219 & -4.3222 & +0.3 & 260   & -0.5574 & -0.5574 &  0.0 \\
    700  & -4.4479 & -4.4477 & -0.2 & 267.7 & -0.5619 & -0.5617 & -0.2 \\
    1000 & -4.6089 & -4.6087 & -0.2 & 275   & -0.56597 & -0.56594 & -0.3 \\
    1300 & -4.7949 & -4.7942 & -0.5 & 282.5 & -0.5706 & -0.5703 & -0.3 \\
    1600 & -4.9991 & -4.9986 & -0.5 & 290   & -0.5747 & -0.5748 & +0.1 \\
    \hline
    \hline
    \end{tabular}
    \label{tab:fe_results}
\end{table*}

\section{Beyond canonical sampling} 

\subsection{Geometry optimization} 

The primary workflow of \textsc{Mlacs} can be effectively repurposed for the exploration of a potential energy surface $V(\Rvec)$, with the specific aim of identifying the minimum energy configuration, thereby functioning as a surrogate optimization algorithm.
In this approach, rather than directly exploring $V(\Rvec)$, we utilize the surrogate potential $\Vmlip(\Rvec)$ to rapidly pinpoint a probable minimum energy configuration.
The identified configuration is then incorporated into the database, enhancing the description of the BO surface in the vicinity of the minimum. The process stops when the last configuration generated complies with some energy, forces and/or stress criterion for the \textit{ab initio} potential.
This algorithm facilitates the efficient optimization of the \textit{ab initio} potential, significantly reducing the number of required energy evaluations.

For the minimization of the surrogate BO surface, \textsc{Mlacs} interfaces with the optimization algorithms implemented in \textsc{Ase} and \textsc{Lammps}.
During the minimization process, as the MLIP improves, the configurations added to the database progressively approach the true minimum energy configuration.
Given that the objective is to locate this configuration, it is crucial to accurately describe these final configurations, as opposed to the initial ones that are farther from the minimum and provide limited information for the optimization procedure.
To address this, we have implemented a weighting strategy where the weight $w_i$ of the configuration $i$ in the fitting process is defined using its index in the database by $w_i \propto i^a$, with $a>0$ a user-defined parameter.
This strategy ensures that configurations closer to the minimum energy configuration have a larger influence on the fit, thereby improving the overall accuracy and efficiency of the optimization process.

\subsection{Transition path sampling} 
\label{subsec:transpath} 
The Nudged Elastic Band (NEB)~\cite{Jonsson_Book_1998, Henkelman_JCP113_2000} method is a very powerful technique to study the minimum energy path (MEP) or transition pathway between two given atomic configurations. This approach is particularly useful for investigating chemical reactions, diffusion processes, and phase transitions. It involves defining initial and final states (e.g., reactants and products in a chemical reaction, or different structural configurations) and intermediate \emph{images} representing the path the system might follow during the transition. A key aspect of the method consists in connecting these images together with elastic springs, before performing an energy minimization of the whole path. This is done by applying a (``nudging'') force to each image, driving it along the adjacent minimum energy configurations. The atomic positions of the images are then iteratively optimized until convergence, typically when the energy difference between two consecutive steps does not exceed a user-defined criterion. 

In the \textsc{Mlacs} package, the NEB method has been adapted for transition path search. In fact, the initial and final states allow us to adjust a first MLIP, which is then employed to search the transition path using the NEB method over a large number of images ($\sim 50$ intermediate images). This calculation provides an initial estimation of the transition path. Along this path, we select an atomic configuration, from which we seek to determine the energy through a DFT calculation. This new configuration is then added to the database, to adjust a new MLIP and compute a new transition path. This process can be repeated several times until convergence of the transition path is achieved. In our case, convergence is considered reached when the energy difference calculated between two iterations over the entire transition path does not exceed a certain tolerance. Once the convergence is achieved, a DFT calculation can be performed for the configuration identified by the MLIP as the highest energy point (\textit{saddle} point) to verify that atomic forces are indeed near zero (at the \textit{saddle} point the energy gradient should be null).

The package includes two distinct implementations of the NEB method, one provided by \textsc{Lammps} and the other by the \textsc{Ase} Python package. Each implementation possesses unique features. For instance, the \textsc{Lammps} implementation does not fix the initial and final images, allowing them to relax into their fundamental states. However, this can be potentially risky in certain cases. On the other hand, the \textsc{Ase} implementation offers the possibility to employ variants of the NEB method, such as the Climbing-Image NEB or the DyNEB (Dynamic NEB), to optimize the search for the saddle point (the point of highest energy). It is important to note that while \textsc{Lammps} can be executed with MPI parallelization over images, this feature is not available for the \textsc{Ase} NEB in the current version of \textsc{Mlacs}.

During the execution of the NEB method within \textsc{Mlacs}, it is necessary to add to the database the energy, forces and stresses corresponding to a new atomic configuration. This configuration can be extracted from the transition path $\xi(\Rvec)$ through two main strategies, depending on whether the goal is to obtain the energy barrier or the shape profile of the transition path. For the former, a new atomic configuration corresponding to a specific value of the reaction coordinate $\xi$ or associated with the saddle point is added to the database. For the latter, in order to sample efficiently the transition path $\xi(\Rvec)$ one can use an atomic configuration in the set $\left \lbrace\textbf{R}_{N} \right \rbrace$ coming from the NEB sampled by the MLIP, either randomly chosen within the spline of the previous NEB path, or resulting from a Bayesian inference using a Gaussian process regression (GPR).

\subsection{Minimum free-energy path}
\label{subsec:constrainDM}
Here, we present a general approach for finding transition paths at finite temperature. In this framework, the free energy changes are computed by integrating the derivatives of the free energy with respect to the reaction coordinate~\cite{Chipot_book_2007, Darve_JCP128_2008, Stoltz_2010_chap3}. 

The computation of free energy, as elucidated in section~\ref{sec:free energy computation}, poses a significant challenge, particularly when the objective is to quantify its variation along a reaction pathway between two low-energy atomic configurations ($A \rightarrow B$). The key challenge of these methods is the definition of the reaction coordinate indexing the transition from one state to another.

One approach to this end is to rely on a path $\Rvec_o(\xi)$, interpolating between the initial state~$A$ and the final one~$B$, and indexed by a scalar parameter~$\xi$. Various strategies can be employed to construct one such path: a linear interpolation of atomic positions, a NEB path based on either 0 K DFT calculations or finite temperature MLIP simulations. 

From this guessed transition path, one can define a reaction coordinate as follows. For a given value~$\xi$ of the scalar parameter, one finds all the atomic configurations which are closest to~$\Rvec_o(\xi)$ than to any other configuration along the path (namely~$\Rvec_o(\xi)'$ for~$\xi' \neq \xi$). These configurations share the same value of the reaction coordinate, and correspond to the hyperplane~$\{\Rvec\}_{\xi}$ defined by the equation:
\begin{equation}\label{eq:defhyperplane}
	\frac{\partial \Rvec_o}{\partial\xi} \cdot \left[\Rvec - \Rvec_o\left( \xi\right) \right]= 0 \mbox{\; .}
\end{equation}
The set $\{\Rvec\}_{\xi}$ can be sampled by a constrained molecular dynamics at a given temperature within this $3N-1$ subspace. The free energy gradient $\partial \Fe /\partial\xi$ is then deduced by averaging the atomic forces coming over~$\{\Rvec\}_{\xi}$~\cite{Swinburne_PRL120_2018, Ciccotti_JCP109_1998}.

 The subsequent procedure involves multiple constrained MD calculations for a set of $\{\xi\}$ values. The free energy gradient is then computed along the complete transition profile $A\rightarrow B$, which determines the minimum free energy path (MFEP):
\begin{eqnarray}\label{eq:intxiF}
	\Delta \Fe (\xi) &=& \int^{\xi}_{A}  \frac{\partial \Fe(\xi')}{\partial\xi} d\xi' \\
	 &=&  \Fe(\xi) - \Fe(A) \mbox{\; .}
\end{eqnarray}

\begin{figure}
\includegraphics[width=\textwidth]{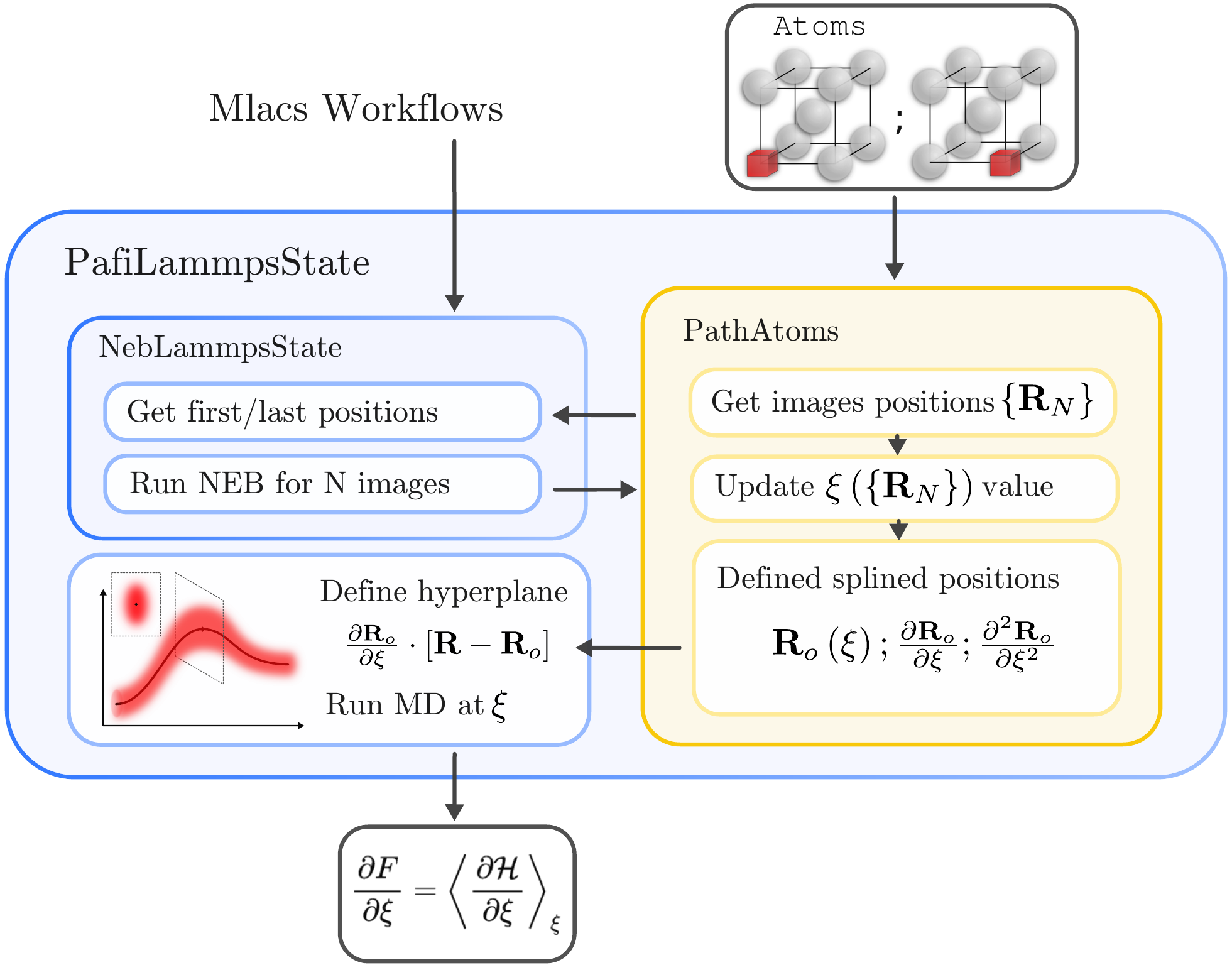}
\caption{\label{fig:MLACS-PAFI} Workflow of the Pafi state object.}
\end{figure}

In \textsc{Mlacs}, constrained sampling (\texttt{PafiLammpsState}, see Fig.~\ref{fig:MLACS-PAFI}) is available within the framework of the PAFI (Projected Average Force Integrator) method~\cite{Swinburne_PRL120_2018} implemented in the \textsc{Lammps} code. 
In addition with the reaction coordinate $\Rvec_o\left( \xi\right)$, the first and second derivatives of positions ($\frac{\partial\Rvec_o}{\partial\xi}$ and $\frac{\partial^{2}\Rvec_o}{\partial\xi^{2}}$) are also required by the \textsc{Lammps} code to initiate the constrained MD (cMD). Subsequently, an atomic configuration is selected from the cMD trajectory.
The forces $\bf{F}_{\xi}$ are then averaged to provide the free energy gradient. The \textsc{Mlacs} workflow is deemed converged when the free energy gradient is minimized below a given criterion. This procedure can be replicated for various reaction coordinates $\xi$ by conducting multiple \textsc{Mlacs} simulations. The complete MFEP is then obtained by integrating the outcomes of all the simulations.

\section{Ecosystem} 

\subsection{Python environment} 

\textsc{Mlacs} is a Python package and most of the simulations are built by running Python scripts to execute simple and independent tasks. Python, a dynamically typed language, offers a clear and concise syntax, making it suitable for creating both simple scripts and complex programs or libraries, such as the now well known Atomic Simulation Environment (\textsc{Ase}) library~\cite{ase-paper}. \textsc{Mlacs} provides multiples interfaces and pre-built objects that can be easily combined to perform complex calculations via scripting or ``on the fly'' using the Python interpreter. This is very convenient for scientific/experimenting approaches (see the example Python \ref{code:mlipfit}). 
\begin{lstlisting}[language=Python, label={code:mlipfit}, caption=Python interpreter example for MLIP fitting.]
>>> from ase.io import read
>>> confs = read('Trajectory.traj', index=':')
>>> 
>>> from mlacs.mlip import SnapDescriptor, LinearPotential
>>> desc = SnapDescriptor(confs[0], 
...                       rcut=6.2,
...                       parameters=dict(twojmax=6))
>>> mlip = LinearPotential(desc)
>>> mlip.update_matrices(confs)
>>> mlip.train_mlip()
\end{lstlisting}
The rising popularity of Python for scientific applications is also largely due to the easy access of free and open-source numerical libraries. \textsc{Mlacs} employs a minimal number of external libraries. The standard version of \textsc{Mlacs} requires only the \textsc{Ase} library and its dependencies. The \textsc{Ase} Python package is designed for setting up, managing, and analyzing atomistic simulations. However, to access all the functionalities of the code, additional packages must be installed. These packages are:
\begin{itemize}
    \item \textsc{Pymbar}, which provides a Python implementation of the MBAR method. This method estimates expectations and free energy differences from equilibrium samples obtained from multiple probability densities~\cite{Shirts2008,shirts2017reweighting},
    \item \textsc{Pyace} coupled with \textsc{Tensorflow}, to fit Atomic Cluster Expansion (ACE) potentials~\cite{ACE_Drautz2019, ACE_Lysogorskiy2021, ACE_Lysogorskiy2023},
    \item \textsc{Scikit-learn}, in order to use Gaussian Process Regressor for transition path sampling, as described in section~\ref{subsec:transpath},
    \item \textsc{NetCDF4}, which is useful for saving thermodynamic properties and structural data in \textsc{NetCDF} format, and for managing the \textsc{Abinit} interface. 
\end{itemize}

\subsection{Additional software} 

\textsc{Mlacs} is also connected with widely recognized software by utilizing compiled executables. The primary one, \textsc{Lammps}, is designed to facilitate effortless developments or expansion with additional features (such as novel force fields like MLIP), resulting in a large community of users and developers. In the \textsc{Mlacs} workflow, \textsc{Lammps} is employed in several ways, particularly for executing molecular dynamics, structural optimization, calculating descriptors, or predicting energies of unknown configurations.

\textsc{Mlacs} is also linked with the \textsc{Mlip-3} code (or the outdated \textsc{Mlip-2})~\cite{Podryabinkin_JCP159_2023}, for training Moment Tensor Potentials (MTP). \textsc{Mlacs} utilizes two executables: \texttt{mlp} for fitting MTP and \texttt{calculate\_efs} for computing energy, forces, and stresses of atomic configurations.

Finally, the \textsc{Mlacs} code is interfaced with an extensive range of electronic structure codes, which serve as the references for the MLIP training process. This is predominantly accomplished through the utilization of the \texttt{Calculator} object supplied by the \textsc{Ase} library. Among the DFT codes compatible with \textsc{Mlacs}, notable ones include \textsc{Vasp}~\cite{Kresse1993,Kresse1996a,Kresse1996b} or \textsc{QuantumEspresso}~\cite{Giannozzi2009,Giannozzi2017}, for instance. The project being originally spearheaded by \textsc{Abinit} developers, \textsc{Mlacs} is specifically designed to operate with \textsc{Abinit}~\cite{Gonze2020,Romero2020}, employing its unique interface \texttt{AbinitManager} to execute calculations and extract output data. Furthermore, \textsc{Mlacs} trajectories can be post-processed using the \textsc{Atdep} tool~\cite{Bottin2020}, incorporated within the \textsc{Abinit} package suite, to obtained thermodynamic properties (the vibrational free energy, phonons, elastic constants, etc $\ldots$). At last, the graphical user interface \textsc{Agate} can be installed to analyze the trajectory saved by \textsc{Mlacs} in the \textsc{NetCDF} format.

\subsection{The MLIP potentials: SNAP, MTP and ACE \label{sec:mlip}} 

While a linear dependence is assumed between the potential energy and the descriptor space, no assumption is made on the form of $\Xmlip(\Rvec)$, enabling the use of a wide variety of descriptors, from a harmonic form to more sophisticated atom centered descriptors such as Atom Centered Symmetry Functions~\cite{Behler2011} or the Smooth Overlap of Atomic Positions~\cite{De2016,Bartk2010}. For the sake of simplicity and robustness, we use the Spectral Neighbor Analysis Potential (SNAP)~\cite{Thompson2015,Wood2018,Cusentino2020} in this work.

To expand the range of configurations that an MLIP can accurately represent within \textsc{Mlacs}, the ACE potential has been implemented using the \textsc{Pyace} module \cite{ACE_Bochkarev2022, ACE_Lysogorskiy2021,  ACE_Ibrahim2023, ACE_Lysogorskiy2023, ACE_Menon2024}. ACE descriptors show excellent transferability and can reliably describe a wide variety of atomic configurations. They utilize products of spherical harmonics to capture interactions up to an arbitrary body order.

The code offers the possibility to use MLIPs in a delta-learning strategy, in which the machine-learning model $\Vmlip(\Rvec)$ is defined as the MLIP potential $\widetilde{\mathbf{D}}(\Rvec){\gamma}$ with a correction coming from a classical potential $\mathcal{V}(\Rvec)$ as:
\begin{equation}
    \Vmlip(\Rvec) = \mathcal{V}(\Rvec) + \widetilde{\mathbf{D}}(\Rvec)\boldsymbol{\gamma} \mbox{\;\; ,}
    \label{eq:deltalearning}
\end{equation}
and the optimization of this potential is performed on the difference between energy, forces and stress of the classical and {\it ab initio} potentials.
Such a strategy can be useful if a fast classical potential is available for the system being studied or to include interactions that cannot be described by the MLIP but for which simple model exists, such as long-range Coulombic interactions. In our implementation, classical potentials available in \textsc{Lammps} can be used, such as Lennard-Jones, EAM, MEAM, Coulomb potential, etc.

\subsection{Classes and objects structures}

\begin{figure}
\includegraphics[width=\textwidth]{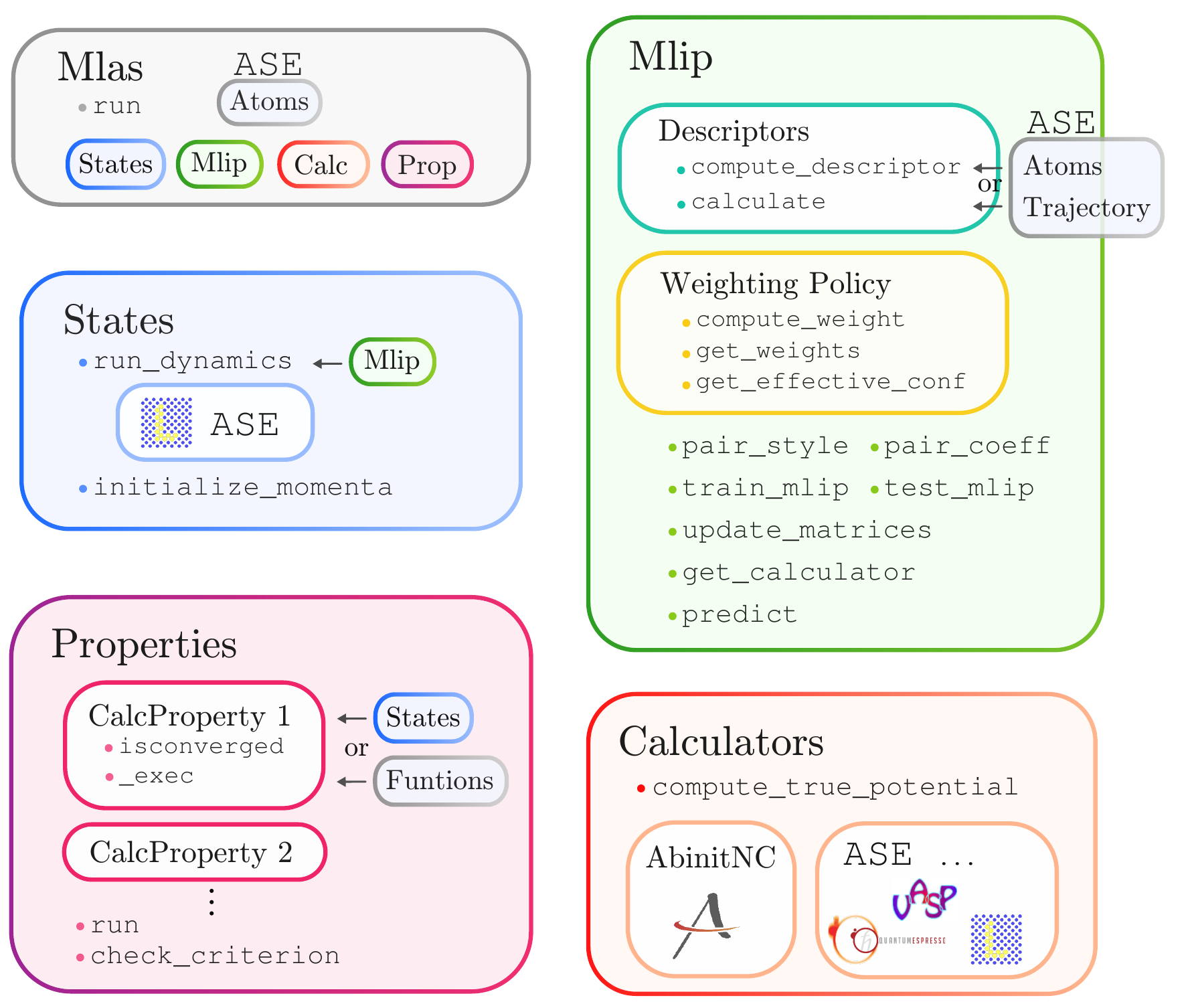}
\caption{\label{fig:MLACS-objects} Scheme of \textsc{Mlacs} main objects and functionalities.}
\end{figure}

The \textsc{Mlacs} library is composed of six primary components. The heart of the entire workflow is carried out by objects of the \texttt{Mlas} type. These ones include the \texttt{OtfMlacs} object, which is primarily used to execute a \textsc{Mlacs} simulation for a specific thermodynamic condition, and the \texttt{MlMinimizer} object, which is utilized for geometry optimization calculations. The main method associated with \texttt{Mlas} objects is the run function, to handle the \texttt{States}, \texttt{Mlip}, \texttt{Calculators}, and the optional \texttt{Properties} classes. All these class objects can be used independently and for specific tasks.

\paragraph{The \texttt{States} classes} They are designed to handle atomic displacements using the MLIP. They are categorized into three types: thermodynamic states for sampling specific thermodynamic states/ensembles (such as NVT, NPT, etc.) via MLMD simulation, ground states for determining/relaxing atomic positions at 0K, and states for sampling transition paths. Each \texttt{States} class follows a consistent structure, featuring a \texttt{run\_dynamics} function to execute the dynamics with the chosen MLIP (refer to the \texttt{States} structure in Fig.~\ref{fig:MLACS-objects}) and an optional \texttt{initialize\_momenta} for setting initial atomic velocities.

\paragraph{The \texttt{Calculators} classes} They conduct reference calculations, primarily utilizing DFT codes. They are typically extensions of \textsc{Ase} \texttt{Calculator} classes. The main function of these classes is \texttt{compute\_true\_potential}, which is responsible for creating, executing, and interpreting the output of DFT calculations. To take advantage of the \textsc{NetCDF} output format, a dedicated interface is employed for DFT calculations using Abinit.

\paragraph{The \texttt{Mlip} classes} They are built to manage all the process of fitting a machine learning potential. This process involves three main steps: first, dimension reduction is performed using descriptors $\Xmlip(\Rvec): \mathbb{R}^{3\Nat}\rightarrow \mathbb{R}^K$; second, a weighting policy $\mathbf{W}_N$ is applied to the database set; and finally, a specified model (primarily linear models in this case) is trained. In the \textsc{Mlacs} package, the process is divided into three primary classes: \texttt{Descriptors}, \texttt{WeightingPolicy}, and the \texttt{Mlip} object that combines the former two. The \texttt{Descriptors} class, associated with an atomic structure (\textsc{Ase} \texttt{Atoms} or \texttt{Trajectory} object), utilizes the \texttt{compute\_descriptor} method to calculate the descriptors. The \texttt{WeightingPolicy} object is responsible for managing weight computations for a particular atomic configuration or the entire database, employing the \texttt{compute\_weight} method. The \texttt{get\_weights} function returns the complete $\mathbf{W}_N$ matrix, while the \texttt{get\_effective\_conf} function provides the actual number of configurations contributing to the database after the weighting policy application. This is mathematically represented as:
    \begin{equation}\label{eq:neff}
        \mathbf{N}_{\text{eff}}^{N} = \frac{\left(\sum_{j=1}^{N} w^N_j \right)^2}{\sum_{j=1}^{N} \left(w^N_j\right)^2} \mbox{\;\;} \in [0,N] \mbox{\;\; .}
    \end{equation}
    The \texttt{Mlip} class subsequently integrates the \texttt{Descriptors} and \texttt{WeightingPolicy} objects to train the model using the \texttt{update\_matrices} and \texttt{train\_mlip} functions. Once trained, the model can predict energies, forces, and stresses of new atomic structures utilizing the \texttt{predict} function, or evaluate a full dataset with the \texttt{test\_mlip} function. The model can be further incorporated into an \textsc{Ase} \texttt{Calculator} object via the \texttt{get\_calculator} method. Alternatively, for \textsc{Lammps} simulations, the \texttt{pair\_style} and \texttt{pair\_coeff} attributes can be considered as inputs using a \texttt{States} object (see example in Python~\ref{code:mlipMD}). 
    \begin{lstlisting}[language=Python, label={code:mlipMD}, caption=Example \ref{code:mlipfit} with a molecular dynamic using the MLIP.]
    >>> from mlacs.state import LammpsState
    >>>
    >>> state = LammpsState(temperature=300, pressure=0)    #NPT
    >>> state.run_dynamics(atoms, 
    ...                    mlip.pair_style, 
    ...                    mlip.pair_coeff)                  
    \end{lstlisting}
    \textsc{Mlacs} is able to manage various types of MLIPs (or descriptor/model combinations), the primary ones being SNAP, MTP, and ACE potentials. These models can also be trained using specific semi-empirical potentials (such as Lennard-Jones, Morse potential, EAM, etc.), where the model prediction corresponds to the difference between the semi-empirical potential and the target one (usually obtained from DFT calculations). This approach is known as ``delta learning'' and is implemented in \textsc{Mlacs} using the \texttt{DeltaLearningPotential} class.

\paragraph{The \texttt{Properties} classes} They manage the measurement of observables. They come in two main flavors of the \texttt{PropertyManager} object: ``custom properties'', which are optional, user-defined properties, and ``routine properties'', which are built-in properties computed systematically without user intervention. Routine properties are typically basic thermodynamic observables (e.g. temperature, volume, stress tensor, etc.) or structural data (cell lengths, velocities, etc.) that are derived from virtually native \textsc{Ase} functions or from the specific choice of \texttt{Calculator}. In contrast, custom properties can be tailored to meet the requirements of each specific problem. Moreover, a convergence criterion can be assigned to a custom property. If all the convergence criteria are met simultaneously, an instruction is passed to the \texttt{Mlas} object to break the main self-consistent cycle. The output values of both property flavors are stored in a \textsc{NetCDF} file that can be either directly accessed in Python, or read by the visualization and post-processing tool \textsc{agate}/\textsc{qAgate}. Finally, information related to the \texttt{WeightingPolicy}, including weights and all weighted properties, is also stored in the \textsc{NetCDF} file.

\paragraph{The \texttt{ThermoState} classes}
They are specialized \texttt{State} classes made to performed thermodynamic integration. Given a set of parameters and a MLIP, they can run all necessary steps: equilibration, the nonequilibrium sampling of the $\lambda$ dependent Hamiltonian introduced in section \ref{sec:free energy computation} and the post-processing.
Three \texttt{ThermoStates} are available: the \texttt{EinsteinSolidState}, which integrates with respect to a system of harmonic oscillators in a solid, the \texttt{UFLiquidState}, for which the integration is done in the liquid state with respect to a Uhlenbeck-Ford potential and the \texttt{ReversibleScalingState} which performs a temperature integration.

\paragraph{The \texttt{ThermodynamicIntegration} classes} This class manages the computation of the free energy, using a list of \texttt{ThermosState} as input and handling their launch and the gathering of post-processed results. A number of instance for each state can be provided, allowing for an estimation of the statistical error.

\subsection{Documentation, Tutorials, Tests and command lines}

A documentation for \textsc{Mlacs} is available at 
\url{https://github.com/mlacs-developers/mlacs/tree/main}, providing information on the theory, installation instruction, tutorials and a complete API.
\textsc{Mlacs} offers user-friendly tutorials in the form of Jupyter notebook codes. These tutorials are specifically designed to run effortlessly on laptop computers by employing the EMT (Effective Medium Theory) potentials from the \textsc{Ase} library for reference, which facilitates rapid evaluation of energies, forces, and stresses. To help users understand the MLIP fitting process, \textsc{Mlacs} includes two tutorials: one for training a SNAP model and another for training MTP models. Furthermore, users can access three tutorials that illustrate how to create standard \textsc{Mlacs} Python inputs for specific applications:
\begin{itemize}
    \item The first tutorial illustrates setting up a \textsc{Mlacs} simulation to sample an FCC bulk of Aluminum in the NPT ensemble at 300~K and 0~GPa.
    \item The second tutorial covers setting up \textsc{Mlacs} with multiple (identical) NPT states and explains how to assign weights to the sampled configurations using MBAR. To showcase the effectiveness of MBAR, we examine a simple Copper crystal at 50~GPa and 400~K, starting from the ground state configuration (0~GPa, 0~K).
    \item The third tutorial presents a method for performing structural optimization on large structures by employing a MLIP as a surrogate model to minimize the number of potential energy calculations.
\end{itemize}
Besides the tutorials, users can probe a variety of examples to draw inspiration from. These examples are located in the \texttt{examples/} directory within the \textsc{Mlacs} package. There is also a lot of integrated tests in \texttt{tests/} directory. Users can execute these tests for verification purposes using the \texttt{pytest} software. 

In addition to the Python library for use in scripts and programs, \textsc{Mlacs} also provides five command-line tools to facilitate post-processing work or data analysis, directly on the command line:
\begin{itemize}
    \item \texttt{mlacs correlation}: plot correlation beetween DFT and MLIP data and compute the root-mean-square error (RMSE). 
    \item \texttt{mlacs plot\_error}: plot the distribution of errors.
    \item \texttt{mlacs plot\_weights}: plot the weight of configurations in the current database and compute the effective number of configurations (see Eq. \ref{eq:neff}).
    \item \texttt{mlacs plot\_neff}: plot the effective number of configurations as a function of the total number of configurations in all preceding database sets, i.e. the \emph{evolution} of $\mathbf{N}_{\text{eff}}^{N}$ as the \textsc{Mlacs} algorithm progresses (cf. Fig.~\ref{fig:neff}).
    \item \texttt{mlacs plot\_thermo}: plot the evolution of some thermodynamic quantities, as presented in Fig.~\ref{fig:mbar}. 
\end{itemize}
All these command line interfaces are designed to easily monitor the \textsc{Mlacs} simulation and the performance of the MLIP fitting. 
Examples of these graphical interfaces are provided below in section~\ref{sec:application}.

\section{Applications}\label{sec:application}
Beyond the initial applications presented in the seminal paper of \textsc{Mlacs}~\cite{Castellano_PRB106_2022} and two recent subsequent studies concerning Au~\cite{Richard2023} and Pu~\cite{Bottin_PRB109_2024}, we would like to highlight several (new) features of \textsc{Mlacs}.

\subsection{The reweighting process using MBAR} \label{subsec:mbar_reweighting}
To illustrate the relevance of MBAR, we study a $4\times 4 \times 4$ supercell of fcc Copper (256 atoms) at $\mathrm{P}=50~\mathrm{GPa}$ and $\mathrm{T}=400~\mathrm{K}$ (cf. second tutorial in the \textsc{Mlacs} package). SNAP descriptors, characterized by cutoff radius $r_{\text{cut}}= 5$~\AA~and maximal angular momentum $j=6$, are used to describe local atomic environments while ensuring dimensionality reduction. Furthermore, to improve the canonical sampling procedure, several independent \texttt{States} are employed, initially drawn from the same thermodynamic ensemble at zero pressure and zero temperature.

\begin{figure}[h!]
\includegraphics[width=\textwidth]{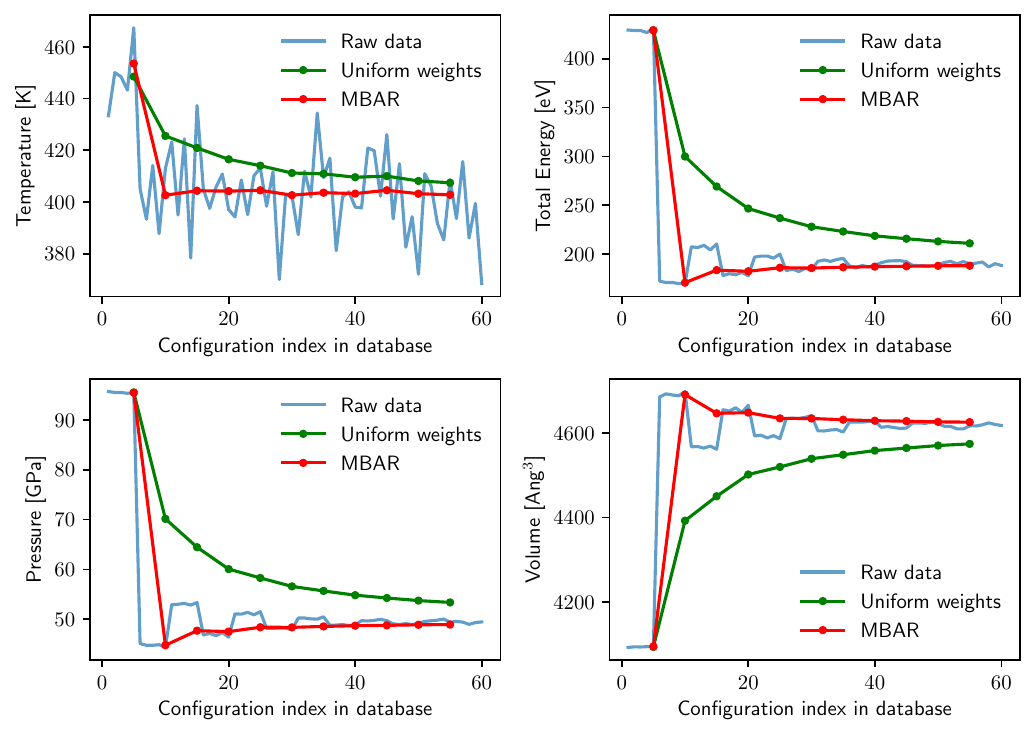}
\caption{\label{fig:mbar} Evolution of key thermodynamic observables during a typical \textsc{Mlacs} run. The system is bulk Cu in the NPT ensemble, as simulated in the second tutorial of the \textsc{Mlacs} package (parameters are detailed in subsection~\ref{subsec:mbar_reweighting}). Temperature and pressure are represented on the left, while the right-hand column shows their extensive counterparts. Here, ensemble averaging is facilitated by collecting results from \emph{five} independent thermodynamic \texttt{States} contributing to a \emph{single} interatomic potential, hence the oscillations in some of the raw data (blue lines). The MBAR results (red dots) are computed at the end of each \textsc{Mlacs} cycle, i.e., every five configurations. As illustrated here, the MBAR strategy converges faster than the naive approach (uniform weights, green dots). This display is generated by the \texttt{mlacs plot\_thermo} command.}
\end{figure}

\begin{figure}[h!]
\includegraphics[width=\textwidth]{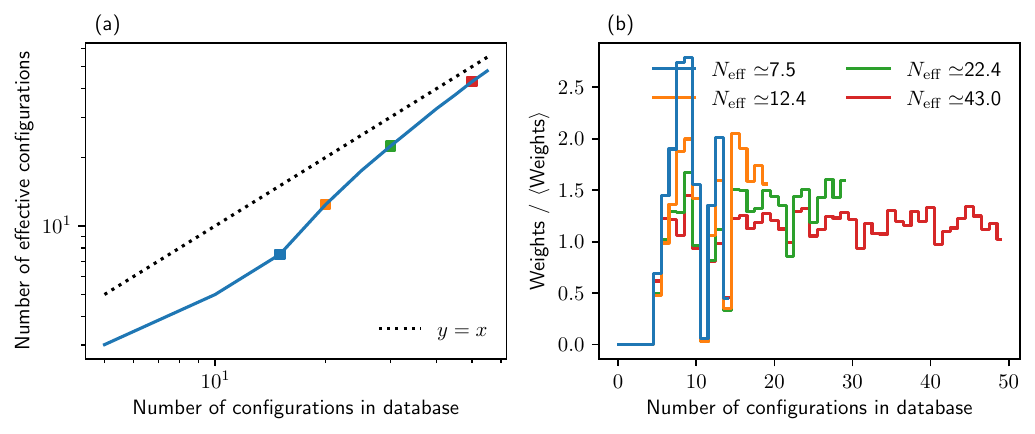}
\caption{\label{fig:neff} (a)~Effective number of configurations $\mathbf{N}_{\text{eff}}^{N}$ as a function of the total number of configurations in database at step $N$, for bulk Cu in the NPT ensemble (cf. subsection~\ref{subsec:mbar_reweighting} and Fig.~\ref{fig:mbar}). The dotted line indicates the ideal case where $\mathbf{N}_{\text{eff}}^{N}$ grows linearly. (b)~Normalized distribution of weights for several numbers of configurations. The set of distributions corresponds to the colored squares in panel (a). These plots are generated by the \texttt{mlacs plot\_neff} command.}
\end{figure}

The evolution of several key thermodynamic observables is presented in Fig.~\ref{fig:mbar}. In particular,  the uniform weighting scheme is directly compared to MBAR, illustrating the faster convergence of the latter. Indeed, a careful analysis of the distributions of weights (cf. Fig.~\ref{fig:neff}) reveals that MBAR typically assigns small (or near zero) weights to the first database configurations, that least represent the target NPT ensemble. Note that, as the number of \textsc{Mlacs} iterations increases and the surrogate interatomic potential approaches its optimum, the distributions shown in panel (b) of Fig.~\ref{fig:neff} become increasingly flat, empirically signalling the convergence of the algorithm.

To further examine the capabilities of MBAR, we summarize the performance of three different approaches in Table~\ref{tab:mbar}. The first calculation involves the previously discussed on-the-fly MBAR reweighting; the second performs a distinct sampling with the uniform weighting scheme; the third scans the second set of atomic configurations and reweighs them \textit{a posteriori} with MBAR. As shown in Table~\ref{tab:mbar}, the MBAR strategies lead to a RMSE on energy one order of magnitude lower than with the uniform weighting policy. Note that the ability of MBAR to be utilized \emph{a posteriori} represents an significant strength and highlights the transferability of \textsc{Mlacs}. Indeed, with this approach, large existing databases can be mined to extract relevant configurations for given new thermodynamic parameters. Finally, Table~\ref{tab:mbar} also demonstrates the efficiency of MBAR schemes when evaluating the free energy differences $\Delta\Fe_{\text{int} \rightarrow \text{AI}}$ (see Eq.~\ref{eq:errorFe}) obtained between the target system ({\it ab initio}) and the system of interest (MLIP). When MBAR is considered, this free energy difference is almost null while a uniform weighting strategy introduces errors.

\begin{table*}[h!]
\centering
	\caption{Various error estimations produced by \textsc{Mlacs} for different weighting policy: (1) MBAR on-the-fly reweighting strategy, (2) uniform weighting policy and (3) \textit{a posteriori} MBAR reweighting strategy. The expression of $\Delta\Fe_{\text{int} \rightarrow \text{AI}}$ is defined by Eq.~\ref{eq:errorFe}.}
\begin{tabular}{lccc} 
    \hline
    \hline
	 & (1) & (2) & (3) \\ 
    \hline
	RMSE Energy (meV/at)   & 0.084 & 3.839 & 0.110 \\
	MAE Energy  (meV/at)   & 0.062 & 2.489 & 0.076 \\
	RMSE Forces (meV/\AA) & 30.46 & 59.35 & 30.20 \\
	MAE Forces  (meV/\AA) & 23.57 & 43.11 & 23.37 \\
	RMSE Stress (MPa)     & 54.38 & 317.79 & 60.33 \\
	MAE Stress  (MPa)     & 39.02 & 208.23 & 43.85 \\
    $\Delta\Fe_{\text{int} \rightarrow \text{AI}}$ (meV/atom) & 0.003 & -0.245 & 0.000 \\ 
    \hline
    \hline
    \end{tabular}
    \label{tab:mbar}
\end{table*}

\subsection{Quality of the canonical sampling: magnesium}

In order to assess the quality of the canonical sample provided by \textsc{Mlacs}, we perform benchmark AIMD calculations for elemental magnesium in the HCP phase at $\mathrm{T}=300~\mathrm{K}$.
We run Langevin dynamics in a $2\times2\times2$ supercell (16 atoms), with 500 MD steps between each DFT step, a time step of 0.5~fs, and a damping parameter of 50~fs.
We employ SNAP descriptors, parameterized with a cutoff radius of 4.2~\AA, and a basis size determined by the maximum angular momentum number $j\!=\!2$.

\begin{figure*}[htb]
\includegraphics[width=\textwidth]{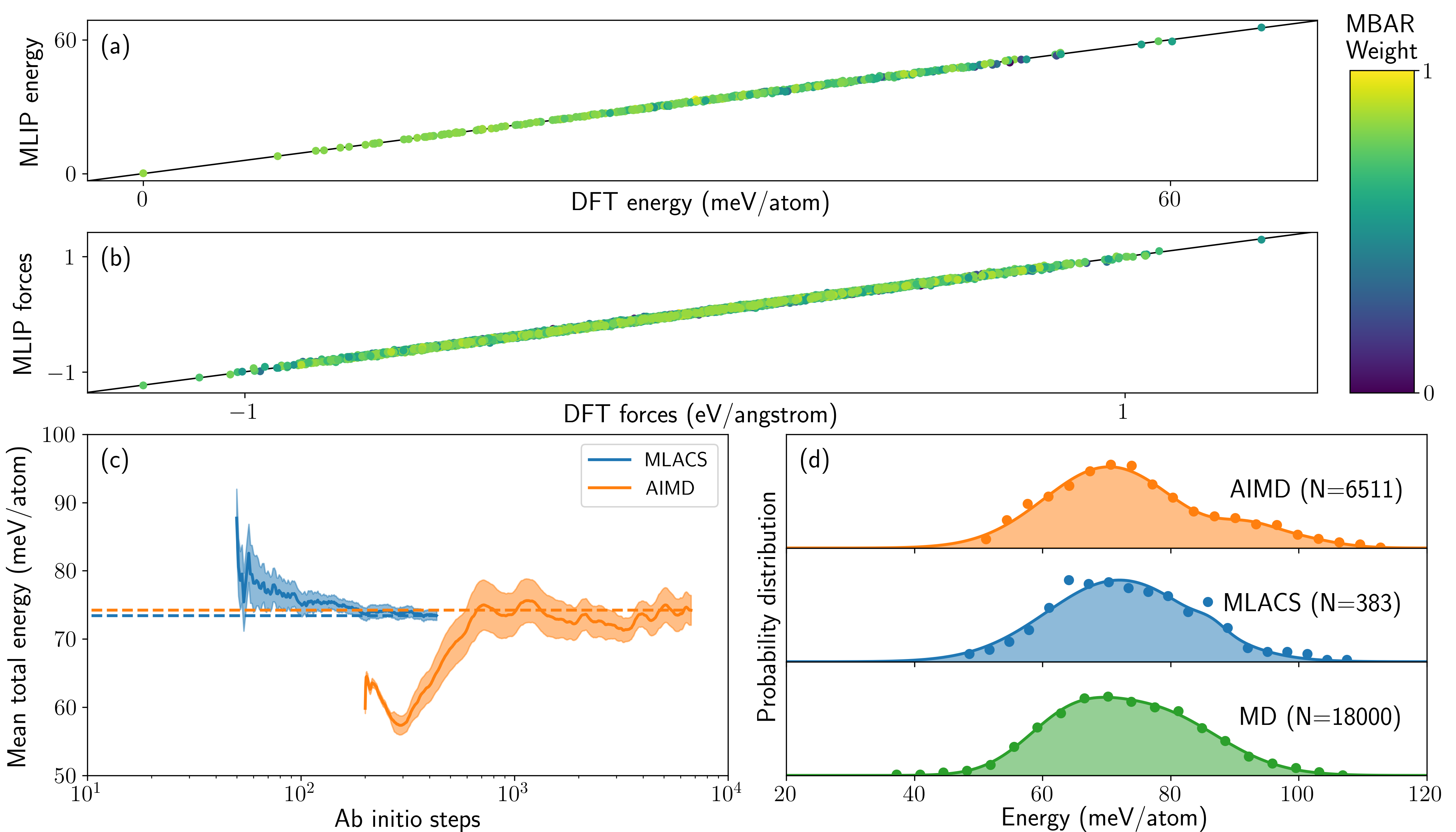}
\caption{\label{fig:Mg_all} 
  AIMD and \textsc{Mlacs} simulations of Mg at $\mathrm{T}=300~\mathrm{K}$.
  (a) 
  Comparison between the surrogate and DFT potential energy,
  with colors showing the MBAR weights.
  (b)
  Comparison between the surrogate and DFT forces.
  (c)
  Convergence of the mean total energy of the supercell
  with the number of ab initio steps.
  The horizontal dashed line marks the value at the end of each simulation,
  and the shaded area represents a confidence interval within one standard deviation.
  (d)
  Normalized probability distribution as a function of total energy
  for three different simulation schemes: AIMD, \textsc{MLACS}, and classical MD.
  $N$ is the number of configurations after thermalization or training steps are removed.
}
\end{figure*}

Figure~\ref{fig:Mg_all} (a-b) presents the potential energy and forces computed with the MLIP against those computed from DFT, along with the weight of the various configurations.
We obtain excellent agreement, with a MAE of 7~meV/\AA \ for the forces, and 0.185 meV/at for the energy.

Figure~\ref{fig:Mg_all} (c) shows the convergence of the mean total energy as a function of the number of DFT steps for both the AIMD and \textsc{Mlacs} calculations.
The mean energy computed by AIMD converges to {74.2~$\pm$~2.2 meV/atom}, while the mean energy computed by \textsc{Mlacs} converges to {73.4~$\pm$~0.7 meV/atom}.
The error that results from sampling the atomic configurations with a surrogate interatomic potential is thus smaller than the statistical uncertainty on the energy.
We note that nearly 10,000 DFT calculations are required to obtain a meaningful sampling for the total energy in AIMD, while the energy converges in less than 200 steps with the \textsc{Mlacs} method.

In Figure~\ref{fig:Mg_all} (d), we show the probability distribution of the configurations as a function of their energy for three different simulation schemes: AIMD, \textsc{MLACS}, and classical MD using the MLIP produced at the end of the \textsc{MLACS} simulation.
We obtain this distribution by collecting the set of visited configurations into a normalized histogram with 20 bins of equal energy width, and fitting a normalized sum of Gaussian functions.
We note that the AIMD probability distribution is bimodal, with a shoulder around 90~meV/atom.
This feature is reproduced by the \textsc{Mlacs} probability distribution, but not by the MD simulation, which displays a unimodal distribution of equivalent width.
Still, the mean total energy computed in MD ({72.9~$\pm$~1.4~meV/atom}) remains accurate within statistical uncertainty.

\subsection{Phase diagram of gold}

\begin{figure*}[htp]
\begin{tabular}{cc}
    \includegraphics[width=0.5\linewidth]{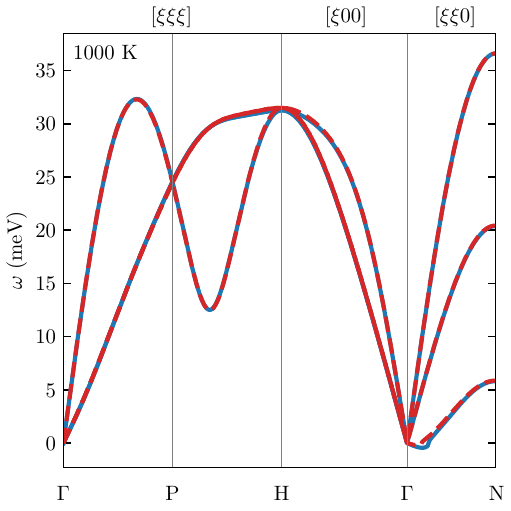} & 
    \includegraphics[width=0.5\linewidth]{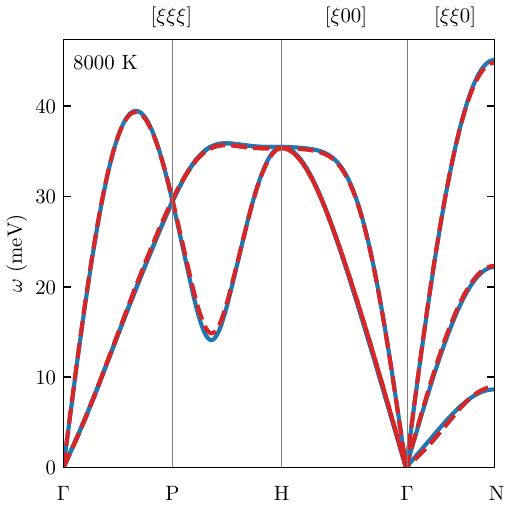}
\end{tabular}    
\caption{\label{fig:bcc_w} Comparisons between \textsc{Mlacs} (solid blue line) and AIMD (dashed red line) bcc-Au phonon spectra. Left pannel: at $\mathrm{V}=12.926~\mathrm{\AA^3/atom}$ and $\mathrm{T}=1000~\mathrm{K}$ (around $\mathrm{P}=100~\mathrm{GPa}$). 
Right panel: at $\mathrm{V}=11.512~\mathrm{\AA^3/atom}$ and $\mathrm{T}=8000~\mathrm{K}$ (around $\mathrm{P}=200~\mathrm{GPa}$).}
\end{figure*}

In the seminal paper~\cite{Castellano_PRB106_2022} of \textsc{Mlacs}, we proved that a surrogate distribution can best approximate the true canonical distribution for many systems. In this example, we consider the bcc phase of gold (a system recently studied using \textsc{Mlacs}, see Ref.~\cite{Richard2023}) and demonstrate the capability of this strategy by comparing AIMD and \textsc{Mlacs} finite temperature phonon frequencies for two thermodynamic conditions. 

DFT calculations are performed using the projector augmented wave method~\cite{Bloch1994,Torrent2008} implemented in \textsc{ABINIT}~\cite{Gonze2020}, the local density approximation for the exchange-correlation functional~\cite{LDA}, a $(2\times2\times2)$ \textbf{k}-point mesh, and on supercells with 128 atoms in the bcc phase. AIMD and \textsc{Mlacs} simulations are carried out in the $NVT$ canonical ensemble, with approximately 5000 times steps and 100 DFT calculations, respectively. 

In Fig.~\ref{fig:bcc_w} we show an excellent agreement between \textsc{Mlacs} and AIMD finite temperature phonon spectra, both at low pressure and temperature (left panel) where the bcc phase is dynamically unstable, and at high pressure and temperature (right panel) where the bcc phase is dynamically stable. This example reveals that \textsc{Mlacs} leads to a huge acceleration (by a factor of 50 approximately) while maintaining an {\it ab initio} accuracy over phonon frequencies, even for anharmonic systems.  

\begin{figure*}[htp]
\begin{tabular}{cc}
     \includegraphics[width=0.49\linewidth]{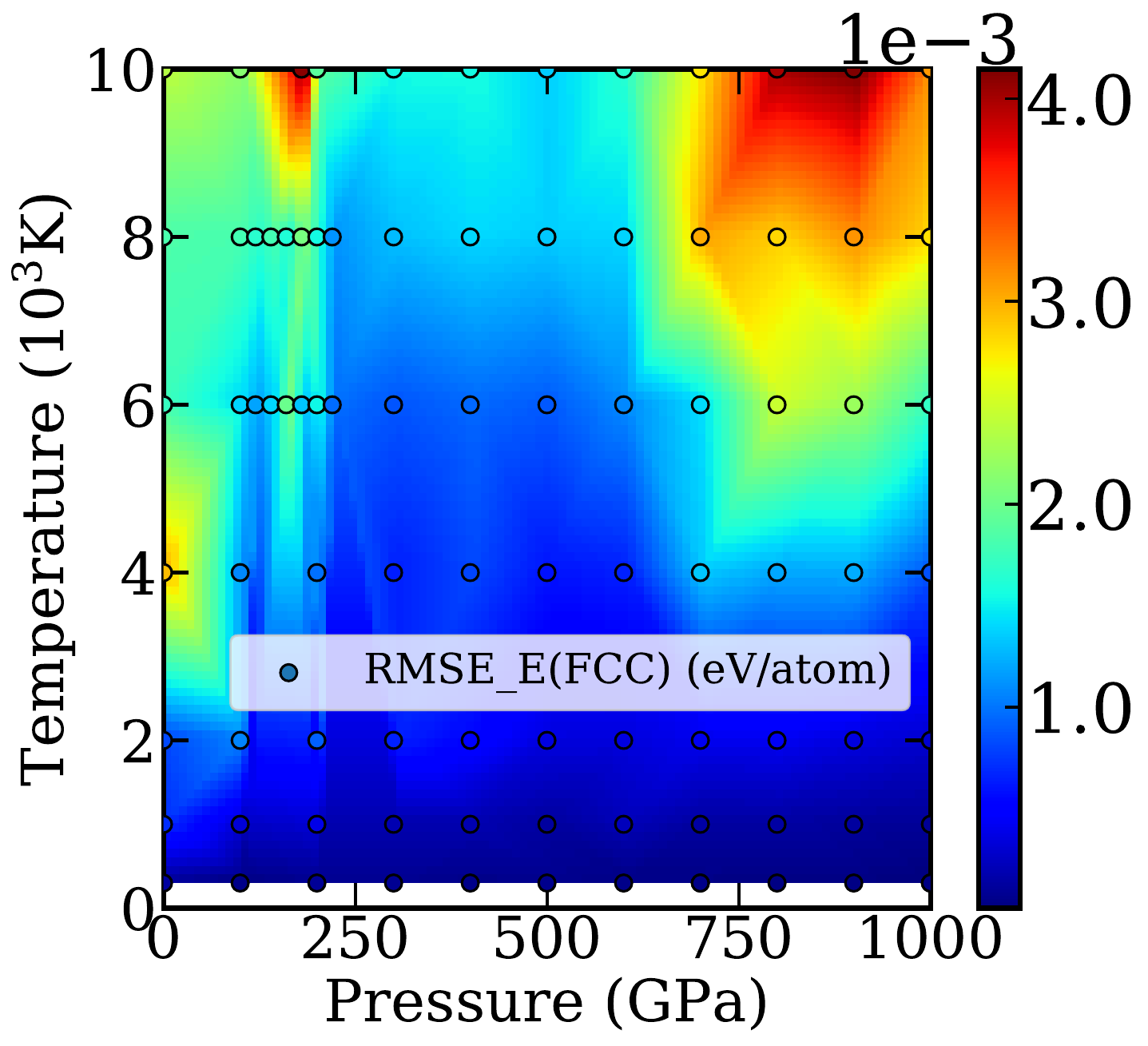} & 
     \includegraphics[width=0.49\linewidth]{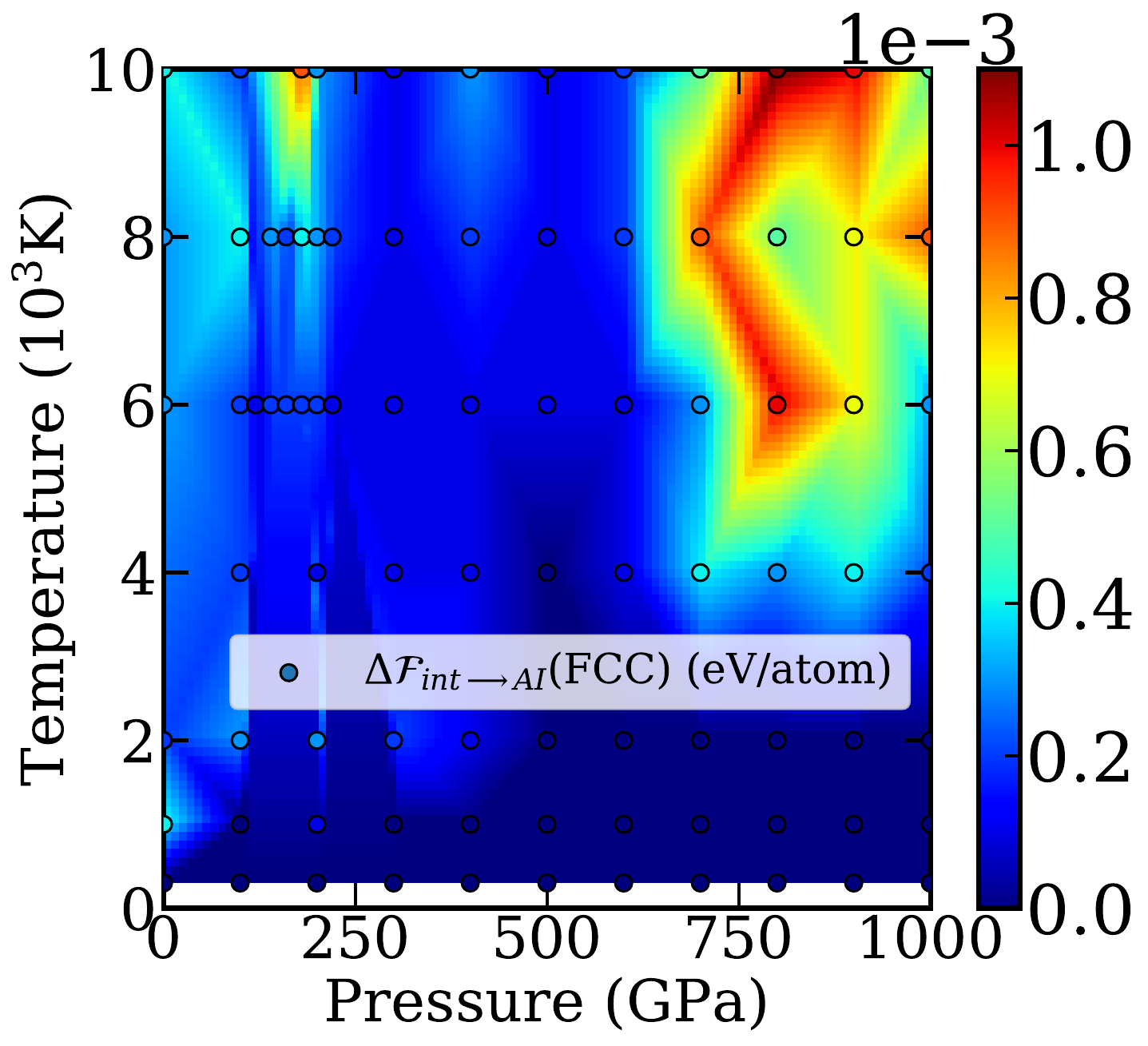}
\end{tabular}    
\caption{\label{fig:deltaf} Errors for the fcc phase of gold on the whole phase diagram. Left panel: the weighted root mean square error on energy. Right panel: free energy error $\Delta \mathcal{F}_{\text{int} \rightarrow \text{AI}}$ computed using the 2$^{nd}$ order cumulant expansion (see Eq.~\ref{eq:errorFe}). Symbols correspond to the values computed for each thermodynamic point whereas the colormap is an interpolation. The color bar is in eV/atom.}
\end{figure*}

To increase the range of thermodynamics conditions over which the comparison is performed, we consider in the following the fcc phase of gold in the [0; 1000] GPa and [0; 10,000] K ranges of pressure and temperature. In the left panel of Fig.~\ref{fig:deltaf}, we show the RMSE between the target DFT and surrogate MLIP energies. The error is far below 1 meV/atom between $\mathrm{T}=300~\mathrm{K}$ and $\mathrm{T}=2000~\mathrm{K}$, is below 2 meV up to $\mathrm{T}=6000~\mathrm{K}$ or $\mathrm{P}=500~\mathrm{GPa}$, and is between 3 and 5~meV for a few thermodynamic states around $\mathrm{T}=10 000~\mathrm{K}$ and $\mathrm{P}=1~\mathrm{TPa}$. 

The above mentioned formalism of NETI is also employed to compute free energies of the various phases of gold (fcc, bcc and hcp) in order to identify different stable crystal structures and to build the phase diagram in this range of temperatures and pressures. While this strategy is out of reach using AIMD simulations for such a high number of thermodynamic points (more than 200), it becomes achievable using classical molecular dynamics simulations with a MLIP potential as performed in this example (see also our previous study in Ref.~\cite{Richard2023}). In the right panel of Fig~\ref{fig:deltaf} we report the free energy error $\Delta \mathcal{F}_{\text{int} \rightarrow \text{AI}}$ between the target DFT and surrogate MLIP systems, computed using the 2$^{nd}$ order cumulant expansion (see Eq.~\ref{eq:errorFe}). The error is about 10$^{-4}$ meV/atom below $\mathrm{T}=2000~\mathrm{K}$, a few 10$^{-4}$ meV/atom below $\mathrm{P}=700~\mathrm{GPa}$ or $\mathrm{T}=6000~\mathrm{K}$, and about 1~meV/atom above. In conclusion, by building a local and optimal MLIP potential for each thermodynamic point, \textsc{Mlacs} demonstrates its ability to maintain an {\it ab initio} accuracy on free energies ($\leq 1$ meV/atom) in a large range of pressures and temperatures. 

\subsection{Geometry optimization of Au\textsubscript{1-x}Cu\textsubscript{x} alloys}

For systems with broken symmetries, such as alloys or crystals with vacancies, structural optimization must be conducted on \emph{large} supercells, containing up to hundreds of atoms. Thus, the minimization procedure requires extensive DFT calculations to converge to the ground state, so that the computational cost can become prohibitive.
To illustrate the acceleration provided by the \textsc{Mlacs} package for these applications, we consider the computation of the excess enthalpy of the Au\textsubscript{1-x}Cu\textsubscript{x} alloy. A key measure of alloy stability,
the excess enthalpy $\Delta H$ is defined as the difference between the alloy ground state energy and the ground state energy of its constituents, i.e., $\Delta H = E(x) - (1 - x) E(0) - x E(1)$, where $x$ denotes the copper concentration.

\begin{figure}[!h]
\includegraphics[width=\textwidth]{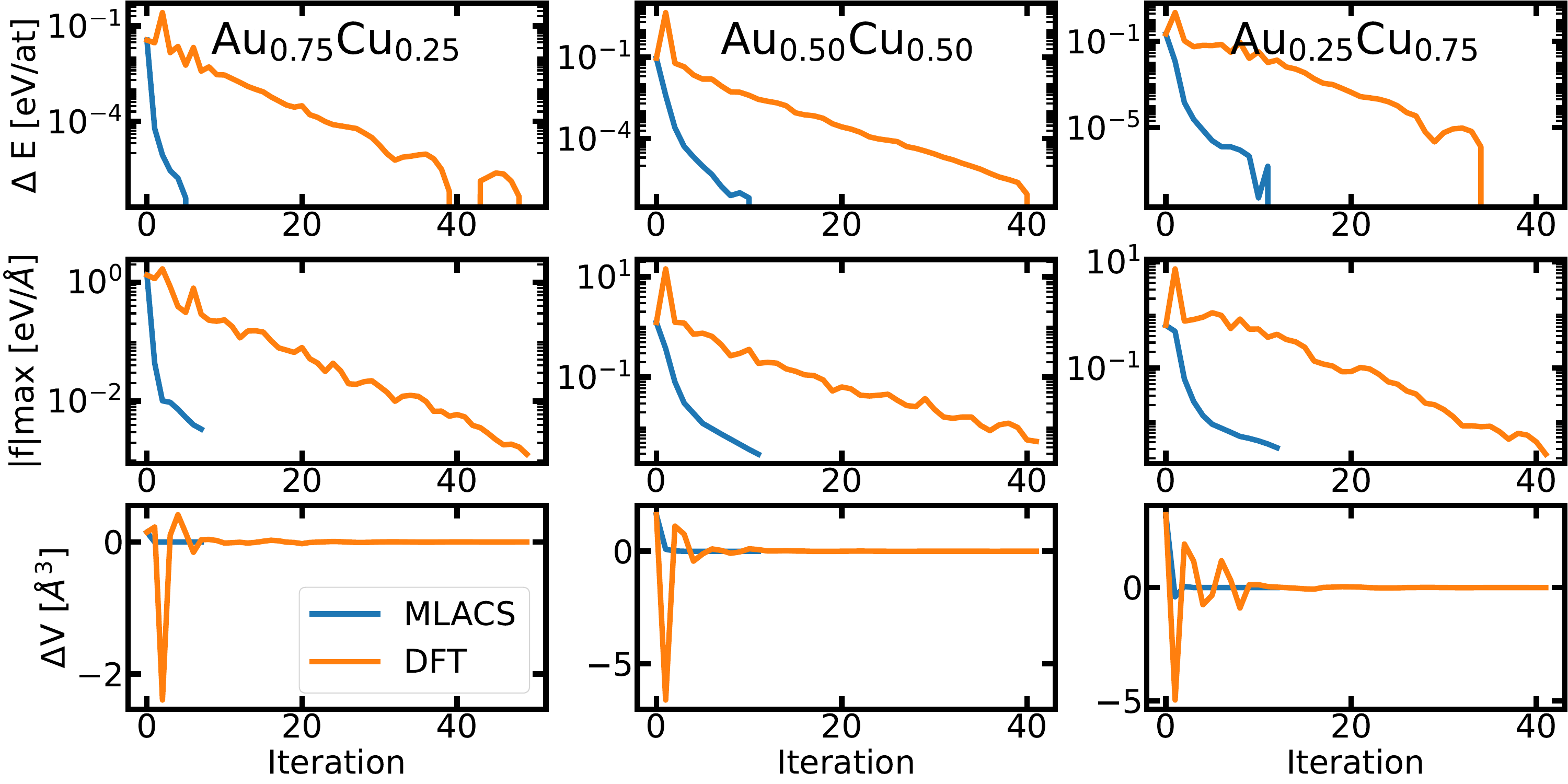}
\caption{\label{fig:AuCuMinimization}
  {Evolution of the energy, maximum absolute forces and volume during the geometry optimization for three representative Au\textsubscript{1-x}Cu\textsubscript{x} alloys, with copper concentrations of $0.25$ (left), $0.5$ (center) and $0.75$ (right).}
  }
\end{figure}

Here, the \textsc{icet} package~\cite{ngqvist2019} was used to generate quasirandom structures~\cite{Zunger1990} derived from a $2\times2\times2$ conventional fcc supercell, at copper concentrations of $0.25$, $0.5$ and $0.75$. For both reference DFT calculations and \textsc{Mlacs}, the optimal structures were obtained by the BFGS algorithm implemented in \textsc{Ase}~\cite{ase-paper}, with convergence criteria of $1$~meV/at, $10$~meV/\AA ~and $0.01$~GPa for energies, forces and stresses, respectively. The evolution of the energy error, maximum absolute forces and volume error are displayed on Fig.~\ref{fig:AuCuMinimization}.
For all three concentrations, the \textsc{Mlacs} algorithm converges about four times faster to the required accuracy than the pure DFT routine.
\begin{figure}[!h]
\includegraphics[width=\textwidth]{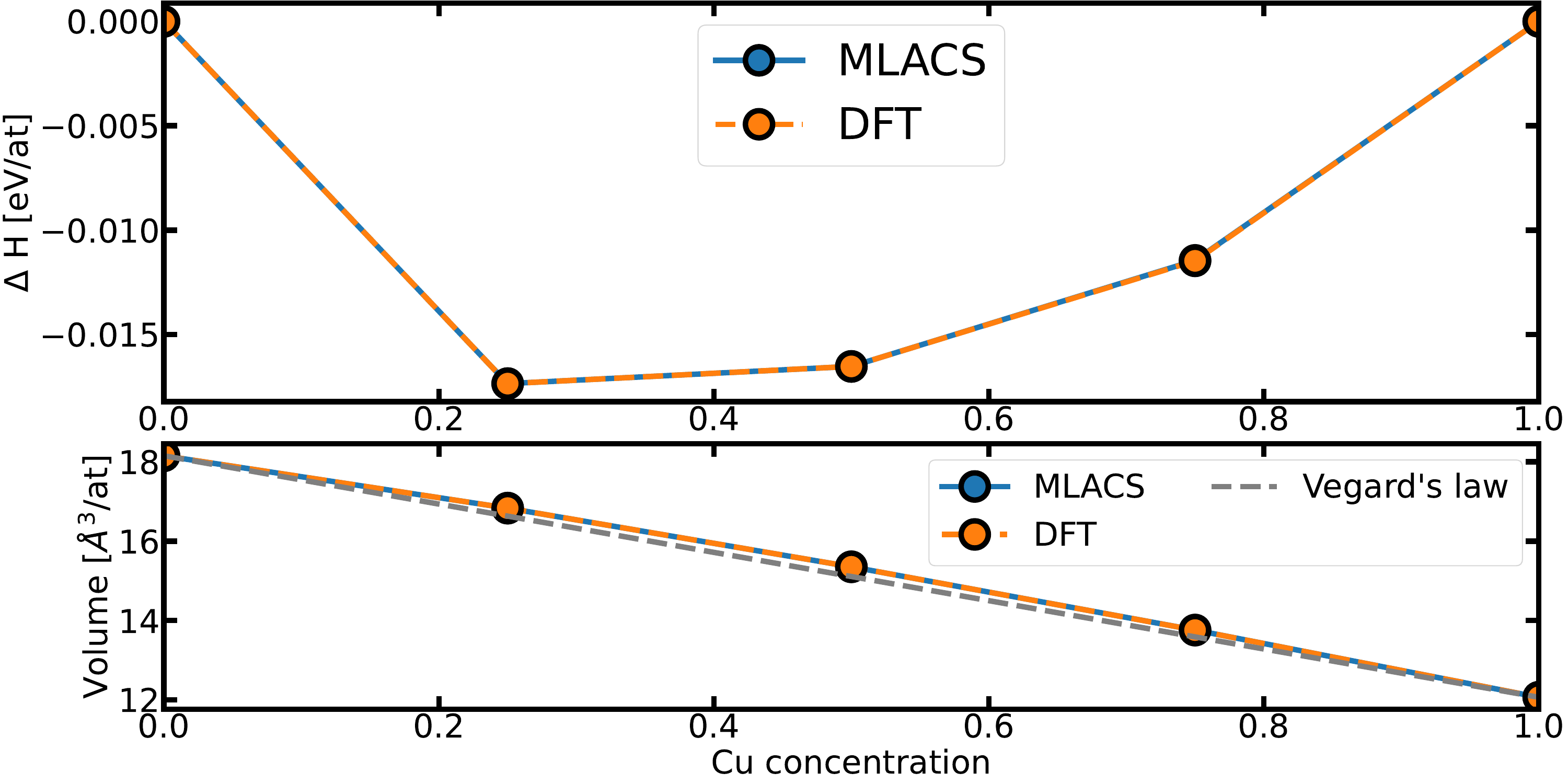}
\caption{\label{fig:AuCuEnthalpie}
  Excess enthalpy (top) and volume (bottom) for Au\textsubscript{1-x}Cu\textsubscript{x} alloy computed with DFT and the \textsc{Mlacs} algorithm.}
\end{figure}

In addition, the upper panel of Fig.~\ref{fig:AuCuEnthalpie} compares the excess enthalpy obtained from \textsc{Mlacs} at several copper concentrations with the reference DFT results. We find an excellent agreement between the two methods, with errors of the order of $10^{-7}$~eV/at.
On the lower panel of Fig.~\ref{fig:AuCuEnthalpie}, the alloy's volume is represented as a function of copper concentration. Interestingly, \textsc{Mlacs} is able to reproduce the subtle deviations from Vegard's law that appear in the DFT reference computation.

In this application, the computational cost is divided by four when using \textsc{Mlacs}, without compromising accuracy. However, for larger, more challenging systems, we expect the computational advantage of \textsc{Mlacs} to be even greater.

\subsection{Diffusion of vacancy in silver} 

\begin{figure}[!h]
\includegraphics[width=\textwidth]{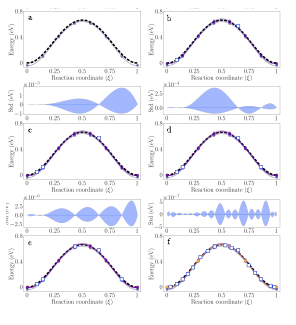}
\caption{\label{fig:AgVacancy}  Nudged Elastic Band method coupled with \textsc{Mlacs} for the study of silver vacancy diffusion, the black dashed curve corresponds to the DFT-NEB calculation (done using the Abinit code). In this case, \textsc{Mlacs} is configured to work with a Gaussian Process Regressor (GPR) to optimize the sampling of the reaction coordinate. Blue filled curves correspond to the standard deviation error computed from the GPR.}
\end{figure}

To enhance the efficiency of DFT calculation, we have also adapted the \textsc{Mlacs} method for transition state sampling. This issue bears a strong resemblance to geometry optimization, involving large cells and broken symmetries, among other complexities. However, the unique requirement here is the consideration of system replicas (refer to subsection~\ref{subsec:transpath}).

To demonstrate the \textsc{Mlacs} acceleration, we consider the case of vacancy diffusion in silver. The migration energy can be readily obtained from the highest energy barrier ($\Delta E_\mathrm{M} = E_{\mathrm{saddle}} - E_{\mathrm{stable}}$) along the reaction coordinate. In this particular instance, we utilize Bayesian inference via the fitting of a Gaussian process regressor (GPR) to identify the most relevant reaction coordinate and efficiently update the database.

Figure~\ref{fig:AgVacancy} presents the initial five steps of the NEB optimization using \textsc{Mlacs}. Post-initialization, a preliminary guess MEP is derived from the MLIP, which is a product of the initialization steps (as shown in Figure~\ref{fig:AgVacancy}.a). Subsequently, two additional configurations are randomly selected along the reaction coordinate, computed using DFT, and incorporated into the database (with the MLIP updated at each step, as depicted in Figure~\ref{fig:AgVacancy}.b). A GPR is then fitted using this database, and the standard deviation error is extracted from the GPR model. The new configuration, corresponding to the reaction coordinate, is chosen based on the maximum error of the model. The database, MLIP, and GPR models are then updated, and this process is iterated multiple times (as illustrated in Figures~\ref{fig:AgVacancy}.c, d, and e) until convergence is achieved (Figure~\ref{fig:AgVacancy}.f).  

\subsection{Liquid water at $\mathrm{T}=400~\mathrm{K}$}
\begin{figure}
    \includegraphics[width=\textwidth]{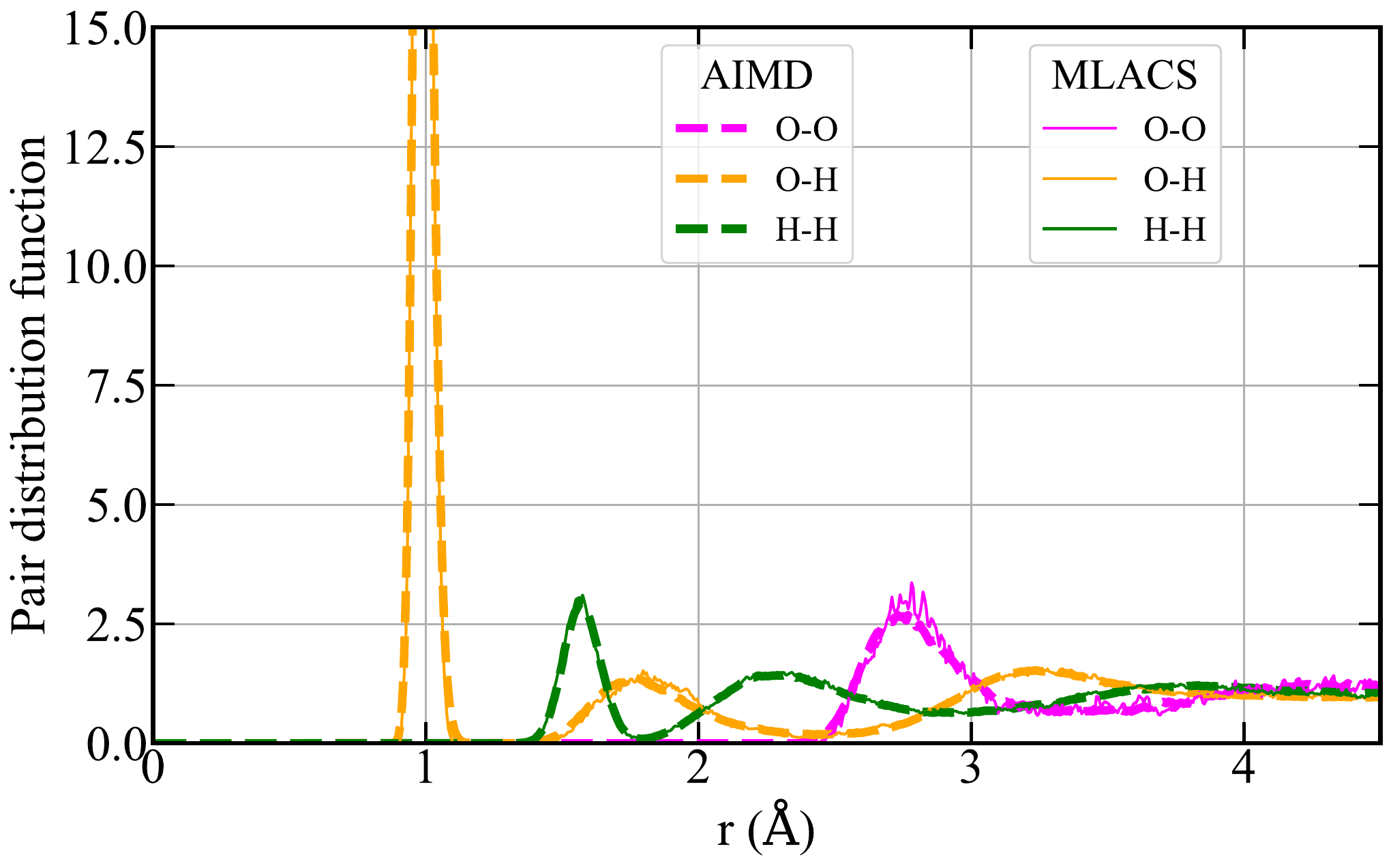}
    \caption{\label{fig:H2O} Pair distribution function of H$_2$O at $\mathrm{T}=400~\mathrm{K}$ obtained using \textsc{Mlacs} and AIMD simulations.}
\end{figure}

Water is a stringent test case for all the ``machine learning" strategies. In this example, we combine the original \textsc{Mlacs} formulation with a delta-learning approach (see Section~\ref{sec:mlip} and Eq.~\ref{eq:deltalearning}). For that purpose, we use a harmonic potential for the O-H bond and a Lennard--Jones plus Coulomb potential for the O-O interactions. Doing that, \textsc{Mlacs} reproduces AIMD results with a good precision (see Fig.~\ref{fig:H2O}). 

The results displayed in Fig.~\ref{fig:H2O} are obtained using AIMD simulations with a supercell of 62 water molecules (186 atoms). The energy cutoff  is set to $24$ Ha, and the \textbf{k}-point mesh is restricted to the $\Gamma$ point. Moreover, a standard Perdew-Burke-Ernzerhof (PBE) exchange and correlation functional is selected, with the D2 van-der-Waals correction proposed by Grimme~\cite{Grimme2006}. Molecular dynamics is performed in the NVT ensemble, resulting in a $10~\mathrm{ps}$ long trajectory with a time step of $0.25~\mathrm{fs}$ (i.e., $4 \times 10^4$ steps). The thermostat temperature is set to $\mathrm{T}=400~\mathrm{K}$, while $\mathrm{P}=0~\mathrm{GPa}$.

In the \textsc{Mlacs} simulation, we used the same computational details as for the AIMD calculations, performed approximately 500 DFT calculations leading to a whole trajectory 15 times longer (150 ps) with 600 000 (SNAP and DFT) time steps. This difference can be at the origin of the small discrepancies obtained between the \textsc{Mlacs} and AIMD pair distribution functions (see Fig.~\ref{fig:H2O}). 

\subsection{Phonon spectrum of UO$_2$ at room temperature} 
\textsc{Mlacs} is especially valuable for systems that require simulating multiple electrons, such as $f$-electron systems\cite{Bottin_PRB109_2024}. A prime example is uranium dioxide (UO$_2$), widely used as nuclear fuel in light water reactors. To model the electronic structure of UO$_2$, we employed the GGA-PBE functional with an added Hubbard-like term. The ($U$, $J$) parameters for uranium cations, similar to those in previous DFT+$U$ studies, are set to $U = 4.5$ eV and $J = 0.54$ eV \cite{dorado2010stability}.
Using this DFT setup, we performed \textsc{Mlacs} calculations at 300 K, generating approximately 100 configurations. The \textsc{Atdep} tool~\cite{Bottin2020} was then used to derive the phonon spectra, converging with around 40 configurations. We estimate that achieving a comparable set of configurations via AIMD would have required at least 4000 time steps, marking an acceleration of nearly two orders of magnitude with \textsc{Mlacs}.

\begin{figure}[h]
\includegraphics[width=\textwidth]{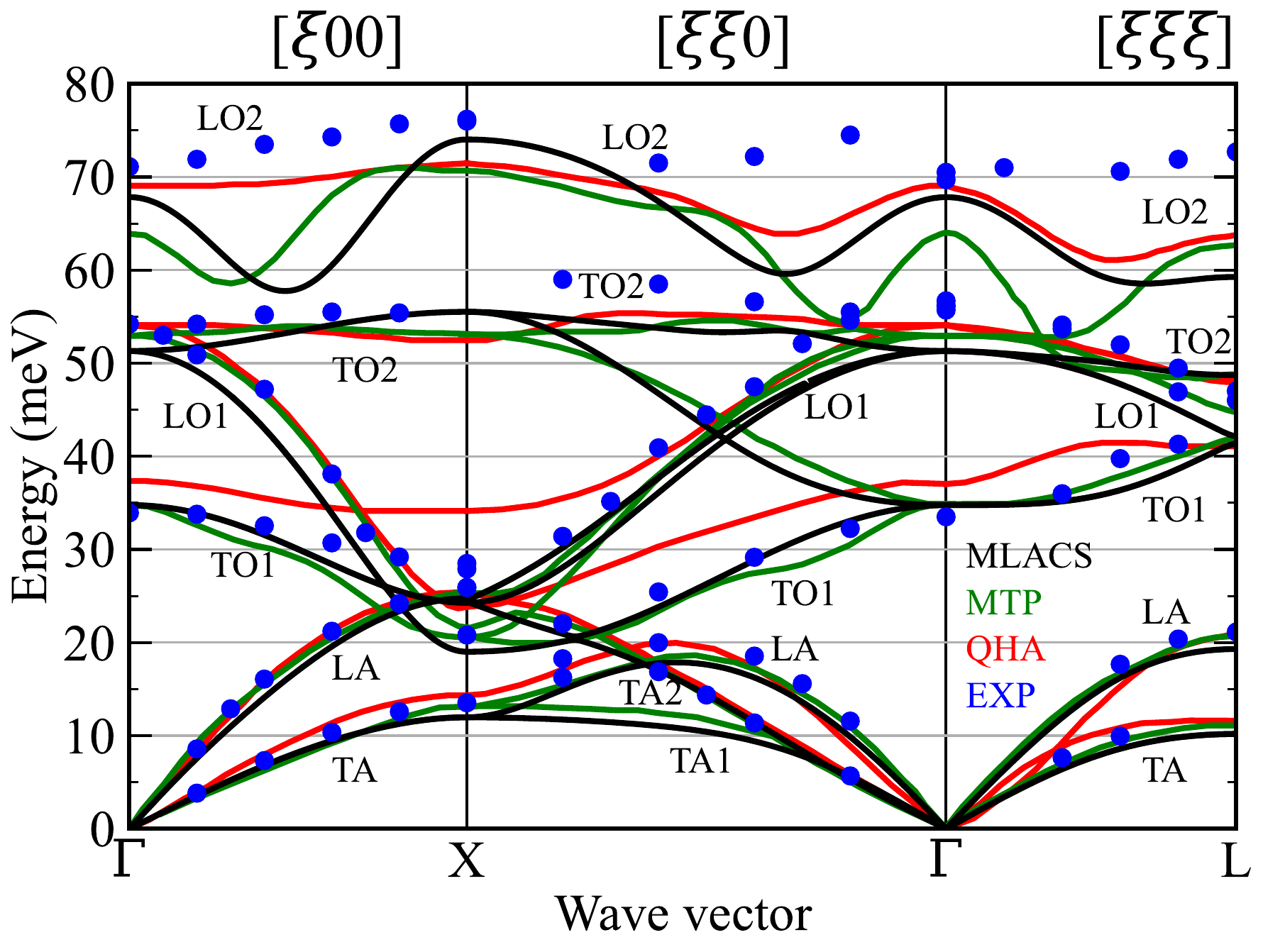}
\caption{\label{fig:phononsThExp}
  Calculated (full lines) and measured (blue circles~\cite{Pang2013}) phonon dispersion relations of UO$_2$ at 300 K along high-symmetry directions in the fcc Brillouin zone. Our results are represented in black, while the quasiharmonic~\cite{Pang2013} and the MTP~\cite{yang22} phonon dispersion are shown in red and green respectively. 
  }
\end{figure}

In Figure \ref{fig:phononsThExp}, we show our \textsc{Mlacs} phonon dispersion relations alongside the experimental measurements (represented as blue circles~\cite{Pang2013}). Additionally, the plot includes the results from the quasi harmonic approximation~\cite{Pang2014} and the MTP potential trained with AIMD~\cite{yang22} in red and green, respectively.
Our calculated phonon frequencies exhibit a strong agreement with the experimental data at high-symmetry points, with notable deviations observed at the L point and for the longitudinal optical branch (LO2), where all calculations display substantial discrepancies ranging between 10-15 meV. Overall, our \textsc{Mlacs} results are comparable to spectra obtained with significantly more computationally demanding methods. 

This demonstrates \textsc{Mlacs} as a promising tool for systematically studying the thermodynamic properties of complex electron systems for which AIMD is often challenging to apply without drastic simplifications to the DFT setup or non-converged calculation parameters aimed at reducing computational time.

\section*{Conclusion}
In this paper, we propose the \textsc{Mlacs} method to accelerate the sampling of a potential energy surface by one to two orders of magnitude compared to AIMD simulations. It uses a machine learning surrogate potential and achieves an accuracy on the order of 1 meV/atom through a self-consistent procedure. The simulated system ultimately remains the {\it ab initio} system, both with respect to the canonical distribution of equilibrium atomic configurations produced by \textsc{Mlacs} and the free energy obtained through thermodynamic integration. 

Through this release, the Python code, its documentation, tutorials, and examples, are made available to the scientific community in a GitHub repository. We believe this advancement could be useful in many studies in materials science and beyond, to accelerate any kind of {\it ab initio} calculations (molecular dynamics, geometry relaxation, minimum energy path, free energy calculations...) that are sometimes computationally too expensive. \textsc{Mlacs} also paves the way for building more efficiently large databases necessary for the development of deep potentials and their use in molecular dynamics~\cite{Zhang_PRL120_2018, Zeng_JCP159_2023}.

\section*{Acknowledgments}
AC acknowledges the Fonds de la Recherche Scientifique (FRS-FNRS Belgium) and Fonds Wetenschappelijk Onderzoek (FWO Belgium) for EOS project CONNECT (Grant No. G.A. 40007563), and Fédération Wallonie Bruxelles and ULiege for
funding ARC project DREAMS (Grant No. G.A. 21/25-11). Simulation time was awarded by PRACE on Discoverer at SofiaTech in Bulgaria (optospin Project No. 2020225411). ON, GG and FB acknowledge GENCI for the access to the
HPC resources of TGCC under the allocation 2024-A0160915134. 
GA and ON acknowledge funding from the Natural Sciences and Engineering Research Council of Canada (NSERC) (Grant No. RGPIN-2019-07149 and DDG-2022-00002), and computational resources from the Digital Research Alliance of Canada and Calcul Québec.

\providecommand{\noopsort}[1]{}\providecommand{\singleletter}[1]{#1}%

\end{document}